\documentstyle[bezier,epsfig,12pt,preprint,tighten,aps]{revtex}
\begin{document}

\draft

\title{\rightline{{\tt (September 1998)}}
\rightline{{\tt UM-P-98/43}}
\rightline{{\tt RCHEP-98/09}}
\ \\
Relic neutrino asymmetry evolution from first principles}
\author{Nicole F. Bell, Raymond R. Volkas and Yvonne Y. Y. Wong}
\address{School of Physics\\
Research Centre for High Energy Physics\\
The University of Melbourne\\
Parkville 3052 Australia\\
(n.bell, r.volkas, y.wong@physics.unimelb.edu.au)}

\maketitle

\begin{abstract}

The exact Quantum Kinetic Equations for a two-flavour active-sterile neutrino system
are used to provide a systematic derivation of approximate evolution equations for
the relic neutrino asymmetry. An extension of the adiabatic approximation for
matter-affected neutrino oscillations is developed which incorporates decoherence due
to collisions. Exact and approximate expressions for the decoherence and repopulation
functions are discussed. A first pass is made over the exact treatment of
multi-flavour partially incoherent oscillations.

\end{abstract}

\newpage

\section{Introduction}

Taken together, the solar neutrino \cite{solar}, atmospheric neutrino \cite{atmos} and
LSND \cite{lsnd} results require at least four light neutrino flavours if neutrino
oscillations are to be their explanation. Consistency with the measured invisible width
of the $Z$ boson then demands that any additional light flavours be sterile with respect
to ordinary weak interactions. With these important experimental results as partial
motivation, the cosmological consequences of sterile neutrinos have recently been
re-examined \cite{prl,ftv,shi,longpaper,astropart,Nnufv,bell,newfoot}. It has been shown
\cite{ftv} that oscillations between active and sterile neutrinos can create significant
asymmetries between the number densities of relic neutrinos and
antineutrinos.\footnote{Synonymns for ``relic neutrino asymmetry'' are:  (i) neutrino
chemical potential (when thermal equilibrium holds), (ii) neutrino degeneracy, and (iii)
nonzero lepton number for the universe.} These asymmetries then have extremely important
consequences for Big Bang Nucleosynthesis (BBN), through their subsequent suppression of
active to sterile neutrino oscillations \cite{longpaper,astropart}, and through the
effect on BBN nuclear reaction rates of a large $\nu_e$ asymmetry \cite{Nnufv,bell}.

Various approximations have been explored in the study of neutrino oscillations in the
early universe, with neutrino asymmetry or lepton number evolution being an important
aspect of these investigations \cite{pioneer,barbieri,mckellar}. The pioneering analyses
neglected the thermal spread in relic neutrino momenta: the evolution of the neutrino
ensemble was assumed to closely track the evolution of neutrinos having the mean
momentum $\langle p \rangle \simeq 3.15 T$. Other approximations often utilised were the
rate equation approximation also neglecting the thermal momentum distribution, or a
Pauli-Boltzmann approach when the momentum spread was not neglected. In the case of
active-sterile oscillations, the number densities of sterile neutrinos were often
neglected if they were small. Pauli blocking, except for the BBN reactions, has usually
been neglected.  Various other more subtle approximations have also been made. For
instance, in the presence of a nonzero neutrino asymmetry, the collision rate for
neutrinos differs from that of antineutrinos. This difference has always been neglected.
The repopulation of active neutrino distributions by weak interactions is another
process often treated approximately.

The main purpose of this paper is to clarify the nature of these various approximations,
and to discuss their regions of applicability more fully and systematically than has
been done in the past. Our approach will take the known exact Quantum Kinetic Equations
(QKEs) \cite{mckellar,otherQKEs} for a two-flavour active-sterile system and then
develop systematic approximations to them, focussing on the evolution of lepton number.
Approximate treatments of lepton number evolution have in the past been vital in
obtaining analytical insight. In fact, the discovery that partially incoherent
active-sterile oscillations can create lepton number exponentially quickly was made
through a treatment that neglected the momentum spread and treated the evolution in the
centre of the MSW resonance very approximately \cite{ftv}. The effect of removing these
approximations was subsequently investigated, and a more exact treatment of lepton
number creation obtained \cite{longpaper}. Nevertheless, the simple and approximate
starting point proved essential in the discovery of this interesting effect, whose
existence had hitherto been overlooked because of the complicated nature of more exact
treatments.  Another important motivation for studying approximation schemes is the
computationally intensive nature of numerical analyses based on the exact QKEs.  The
neutrino momentum spread is computationally demanding, especially during the BBN epoch
when neutrinos are out of thermal equilibrium.

Another goal of this paper is to address multi-flavour effects in the early universe
from first principles for the first time. Previous analyses have either considered a
two-flavour system in
isolation, or have approximated the multi-flavour system via pairwise two-flavour
subsystems \cite{longpaper,astropart,newfoot}. (Note that some genuine three-flavour
effects were considered for fully coherent oscillations in Ref.\cite{bell}.)
We will clarify how the ``pairwise two-flavour
approximation'' arises from a more exact treatment. The three-flavour
problem is particularly important, because one of the principal applications of
lepton number creation lies in the suppression of the large angle $\nu_{\mu} - \nu_s$
oscillations that may solve the atmospheric neutrino anomaly \cite{numunus}. The lepton
number
responsible for this suppression is quite possibly or plausibly created by small
angle $\nu_\tau - \nu_s$ oscillations \cite{longpaper,astropart}. If so, then a full
three-flavour analysis of the $\nu_{\mu}$, $\nu_{\tau}$ and $\nu_s$ system is warranted.

This paper is structured as follows: Section \ref{main1} discusses exact and
approximate treatments of the evolution of lepton number in the context of a simple
two-flavour active-sterile system. Section \ref{mirror} briefly discusses the
important case where the strictly sterile neutrino is replaced by a mirror neutrino.
Section \ref{main2} is devoted to a study of multi-flavour effects, and
Section \ref{conc} is a conclusion.

\section{The two-flavour active-sterile neutrino system: from first principles to
useful approximations}
\label{main1}

The early universe contains, in part, an ensemble of neutrinos.\footnote{When there
is no chance of confusion, or when the distinction is unimportant, we will use the
term ``neutrino'' to mean neutrino and/or antineutrino, active and/or sterile. When
the distinction is important, we will specify which one we mean.} Their evolution in
time is affected by three phenomena. The first is simply the expansion of the
universe. The second is decohering collisions with the background medium. The third
is, of course, the coherent process of oscillations governed by a matter-dependent
effective Hamiltonian. The effect of collisions and oscillations is quantified using
the Quantum Kinetic Equations. The QKEs generalise the Pauli-Boltzmann Equations to include
quantum coherence between the particle species, in our case the active and sterile
neutrinos (and active and sterile antineutrinos).

We will focus on the two-flavour system comprising one active and one sterile neutrino
species in this section. We do so because two-flavour active-sterile oscillations lead
to the rapid creation of lepton number when the oscillation parameters are in the
correct region. The multi-flavour case is deferred to Section \ref{main2}. Our task is to
track the evolution of the neutrino and antineutrino ensemble, focusing in particular
on the evolution of lepton number. We will take as our initial, high temperature,
condition that the number densities of sterile neutrinos and antineutrinos are zero.
This is expected simply because, by definition, sterile neutrinos do not feel
electroweak (or strong!) interactions and so will decouple very
early.\footnote{Negligible initial sterile neutrino number densities is a consistent
assumption because the neutrino system is frozen at high temperatures --- see later. It
is at lower temperatures that sterile neutrino production due to partially incoherent
active-sterile oscillations threaten overproduction.} We will focus on the epoch between
$T = m_{\mu} \simeq 100\ \text{MeV}$ and BBN ($T \sim 0.1 - 1\ \text{MeV}$), so that the
background plasma consists of essentially only photons, electrons, positrons, neutrinos,
antineutrinos and whatever amount of sterile neutrinos and antineutrinos get generated
through oscillations.

\subsection{The exact Quantum Kinetic Equations}

We first write down and discuss the exact QKEs, partly by way of review and to define
notation, and partly to discuss approximations to the decoherence and repopulation
functions (to be defined shortly).

Consider a two-flavour active-sterile system composed of $\nu_{\alpha}$ and
$\nu_s$, where $\alpha$ denotes either $e$, $\mu$ or $\tau$. The 1-body reduced
density
matrices describing $\nu_{\alpha} - \nu_s$ and $\overline{\nu}_{\alpha} -
\overline{\nu}_s$ oscillations are given, respectively, by
\begin{eqnarray}
\rho(p) & = & \frac{1}{2}P_0(p)[1 + {\bf P}(p) \cdot {\bf \sigma}],\ \\
\overline{\rho}(p) & = & \frac{1}{2}\overline{P}_0(p)[1 + \overline{{\bf P}}(p)
\cdot {\bf \sigma}],
\end{eqnarray}
where
\begin{equation}
{\bf P}(p) \equiv P_x(p)
\hat{x} + P_y(p) \hat{y} + P_z(p) \hat{z},
\end{equation}
plus an analogous
expression for antineutrinos. The functions $P_0$ and ${\bf P}$ are just the coefficients of
a convenient expansion of $\rho$ in terms of Pauli matrices ${\bf \sigma}$ and the identity
matrix (${\bf P}$ is often called the ``polarisation''). The quantity $p$ is the
neutrino or antineutrino momentum. It will be understood that
$P_0$ and ${\bf P}$ depend on time (or, equivalently, temperature via $t \simeq
m_{\text{Planck}}/11T^2$) as well as momentum. The diagonal entries of $\rho$
($\overline{\rho}$) are relative number density distributions in momentum space of
$\nu_{\alpha}$ ($\overline{\nu}_{\alpha}$) and $\nu_s$
($\overline{\nu}_s$):
\begin{eqnarray}
N_{\alpha}(p) & = & \frac{1}{2} P_0(p)
[1 + P_z(p)] N^{\text{eq}}(p,0),\ \\
N_s(p) & = & \frac{1}{2} P_0(p) [1 -
P_z(p)] N^{\text{eq}}(p,0), \label{relnums}
\end{eqnarray}
where $N^{\text{eq}}(p,0)$ is
a reference number density distribution that we have set equal to the equilibrium
Fermi-Dirac distribution,
\begin{equation}
N^{\text{eq}}(p,\mu) = \frac{1}{2\pi^2}\frac{p^2}{1 + \exp\left(\frac{p-\mu}{T}\right)},
\end{equation}
with zero chemical potential.
This choice amounts to normalising $\rho$ such that $\text{Tr}(\rho) =
2$ if thermal equilibrium holds and the chemical potential is zero. The antineutrino
relative number densities are given similarly. Since the physical interpretation of
$\rho$ is related to number density distribution {\it ratios}, the expansion of the
universe does not directly contribute to the time evolution of $\rho$. Note that
$N^{\text{eq}}(p,0)$ depends on time only through the expansion of the universe, and
not on coherent or incoherent matter effects.

The evolution of $P_0(p)$ and ${\bf P}(p)$ is governed by the QKEs
\begin{eqnarray}
\frac{\partial}{\partial t}{\bf P}(p) & = & {\bf V}(p) \times {\bf P}(p) + [1 -
P_z(p)] \left[\frac{\partial}{\partial t}\ln P_0(p)\right] \hat{z}\nonumber \\
& - & \left[D(p) + \frac{\partial}{\partial t}\ln P_0(p)\right]
\left[P_x(p)\hat{x} + P_y(p)\hat{y}\right],
\label{vecPeqn}
\end{eqnarray}
and
\begin{equation}
\frac{\partial}{\partial t}P_0(p) = R(p).
\label{P0eqn}
\end{equation}
These equations are obtained by evolving the full density matrix for all particles
in the
plasma forward in time by means of the $S$ matrix, and then tracing over all degrees of
freedom other than $\nu_{\alpha}$ and $\nu_s$. The essential difference between the
Quantum Kinetic and Pauli-Boltzmann approaches is that the former time evolves amplitudes while
the latter evolves probabilities. See, for example, Ref.\cite{mckellar} for a
detailed derivation of the QKEs and further discussion.
The antineutrino QKEs take an identical form, but with the substitutions
\begin{equation}
P_0 \to \overline{P}_0,\ \ {\bf P} \to \overline{\bf P},\ \ {\bf V} \to
\overline{{\bf V}},\ \ D \to \overline{D},\ \ R \to \overline{R}.
\end{equation}
We now discuss the form and significance of each of the terms above.\footnote{These
equations are exact for strictly sterile species. All contributions due to possible
right-handed weak interactions and the like have been neglected.}

The function ${\bf V}(p)$ describes the quantally coherent part of the
matter-affected
evolution of the $\nu_{\alpha} - \nu_s$ subsystem. It is given by
\begin{equation}
{\bf V}(p) = \beta(p)\hat{x} + \lambda(p)\hat{z},
\end{equation}
with
\begin{eqnarray}
\beta(p) & = & \frac{\Delta m^2}{2p} \sin2\theta_0,\\
\lambda(p) & = & - \frac{\Delta m^2}{2p}\cos2\theta_0 + V_{\alpha}.
\end{eqnarray}
The quantities
$\Delta m^2$ and $\theta_0$ are the mass-squared difference and vacuum mixing angle
for $\nu_{\alpha} - \nu_s$ oscillations, respectively. We define
the mass eigenstate neutrinos $\nu_{a,b}$ by
\begin{equation}
\nu_{\alpha} = \cos\theta_0\nu_a + \sin\theta_0\nu_b,\quad
\nu_s = -\sin\theta_0\nu_a + \cos\theta_0\nu_b,
\end{equation}
with $\theta_0$ defined so that $\cos2\theta_0 \ge 0$ and $\Delta m^2 \equiv m_b^2 -
m_a^2$. The function $V_{\alpha}$ is the effective matter potential \cite{wolfenstein}.
To leading
order, including the lowest order finite temperature term, it is given by
\cite{Veffcalc}
\begin{equation}
V_{\alpha} = \frac{\Delta m^2}{2p}[- a(p) + b(p)],
\label{Valpha}
\end{equation}
with the dimensionless functions $a(p)$ and $b(p)$ given by
\begin{eqnarray}
a(p) & = & - \frac{4\zeta(3)\sqrt{2}}{\pi^2}\frac{G_F T^3 p}{\Delta m^2}
L^{(\alpha)},\\
b(p) & = & - \frac{4\zeta(3)\sqrt{2} A_{\alpha}}{\pi^2} \frac{G_F T^4
p^2}{\Delta m^2 m_W^2},
\end{eqnarray}
where $T$ is temperature, $G_F$ is the Fermi constant, $m_W$ is the $W$-boson
mass, $\zeta$ is the Riemann zeta
function, $A_e \simeq 17$, $A_{\mu,\tau} \simeq 4.9$  and
\begin{equation}
L^{(\alpha)} = L_{\alpha} + L_{e} + L_{\mu} + L_{\tau} + \eta.
\label{L}
\end{equation}
The quantities on the righthand side of this equation are the various asymmetries, given
by
\begin{equation}
L_f = \frac{n_f - n_{\overline{f}}}{n_{\gamma}},
\label{asymdef}
\end{equation}
where
\begin{equation}
n_f = \int N_f(p) dp
\label{Nn}
\end{equation}
are number densities (note that we will abbreviate $L_{\nu_{\alpha}}$ by $L_{\alpha}$
in this Section). The $a(p)$ term in Eq.(\ref{Valpha}) is the leading order
density-dependent (Wolfenstein) contribution to the effective potential, while the $b(p)$
term is the leading order finite temperature term. (Note that the Wolfenstein term also
depends on temperature, but only indirectly through the cosmological red-shifting of fermion
number density distributions.) Observe that ${\bf V}$ depends on $\rho$ through the
dependence of $a(p)$ on $L_{\alpha}$. As we discuss later, this makes the evolution of
$L_{\alpha}$ non-linear.
It is important to notice that the dependence of ${\bf V}$ on $L_{\alpha}$ is an $O(G_F)$
effect (due to coherent forward scattering), rather than an $O(G_F^2)$ effect. The quantity
$\eta$ is a small
term related to the cosmological baryon-antibaryon asymmetry. For antineutrinos, the
corresponding function $\overline{{\bf V}}$ is obtained from ${\bf V}$ by setting
\begin{equation}
\overline{\lambda}(p) = - \frac{\Delta m^2}{2p}[\cos 2\theta_0 - b(p) - a(p)],
\end{equation}
and replacing $\lambda$ by $\overline{\lambda}$. In other words, one simply replaces
$L^{(\alpha)}$ by $-L^{(\alpha)}$.

The Mikheyev-Smirnov-Wolfenstein (MSW) \cite{wolfenstein,ms} resonance conditions are
\begin{equation}
\cos 2\theta_0 + a(p) - b(p)  = 0,\quad \cos 2\theta_0 - a(p) - b(p) = 0,
\label{res}
\end{equation}
for neutrinos and antineutrinos, respectively. Note that prior to the creation of lepton
number, both
resonance conditions are simply $b(p) = \cos 2\theta_0$, which can only be satisfied if
\begin{equation}
\Delta m^2 < 0.
\label{Deltamneg}
\end{equation}
This condition plays an important role (see later). It is also important to realise that
the resonance conditions of Eq.(\ref{res}) at a given temperature $T$ are met only for
neutrinos and antineutrinos of
a certain momentum $p = p_{\text{res}}$. This is why the spread of neutrino momenta plays a
significant role in the analysis \cite{longpaper}.

The {\it decoherence} or {\it damping function} $D(p)$ quantifies the loss of quantal
coherence due to $\nu_{\alpha}$ collisions with
the background medium. Its exact expression is derived in a general form in
Ref.\cite{mckellar}, and is given in the Appendix by Eq.(\ref{eq:D}). When thermal
equilibrium holds, the general expression for $D$ reduces to a useful compact form,
given by
\begin{equation}
D(p) = \frac{1}{2} \Gamma_{\alpha}(p)
\end{equation}
where $\Gamma_{\alpha}(p)$ is the total collision rate for $\nu_{\alpha}$'s with
momentum $p$. By examining the momentum dependence of the specific weak collision processes
operating between the neutrino decoupling temperature and $100$ MeV, we see that
\begin{equation}
\Gamma_{\alpha}(p) = \langle \Gamma_{\alpha}\rangle \frac{p}{\langle p \rangle_0},
\label{Gammap}
\end{equation}
where
\begin{equation}
\langle p \rangle_0  = \frac{7\pi^4}{180\zeta(3)}T \simeq 3.15 T
\end{equation}
is the average momentum for a relativistic Fermi-Dirac distribution with zero chemical
potential (indicated by the subscript ``$0$''), and $\langle \Gamma_{\alpha}\rangle$
is the thermally averaged total collision rate for $\nu_{\alpha}$. This last quantity can be
expanded as a power series in $L_{\alpha}$,
\begin{equation}
\langle\Gamma_{\alpha}\rangle = G_F^2 T^5 y_{\alpha}(1 - z_{\alpha} L_{\alpha}) +
O(L^2_{\alpha}),
\label{explicitGammap}
\end{equation}
with $y_e \simeq 4$, $y_{\mu,\tau} \simeq 2.9$, $z_e \simeq 0.1$ and $z_{\mu,\tau}
\simeq 0.04$. The $y_{\alpha}$ term,
\begin{equation}
\label{Gammma_ave}
\langle\Gamma_{\alpha}\rangle_0 \equiv y_{\alpha} G^2_F T^5,
\end{equation}
is the thermally averaged collision rate at zero chemical potential, while the $z_{\alpha}$
term is the first order
correction due to a finite neutrino asymmetry. For antineutrinos, the corresponding
expression is
\begin{equation}
\overline{D}(p) = \frac{1}{2}\overline{\Gamma}_{\alpha}(p) =
\frac{1}{2} G_F^2 T^5 y_{\alpha}\frac{p}{\langle p \rangle_0}(1 + z_{\alpha} L_{\alpha}) +
O(L^2_{\alpha}),
\label{explicitGammabarp}.
\end{equation}
In previous studies \cite{longpaper,astropart}, Eq.(\ref{Gammap}) was adopted as an
approximate result for
$\Gamma_{\alpha}(p)$. Actually, it turns out to be {\it exact}, as we show in the
Appendix.

It is important to observe that although $D$ is in principle a dynamical quantity through
its dependence on neutrino number densities and hence on $\rho$ [see Eq.(\ref{eq:D})], it
becomes approximately
kinematic provided two conditions hold: (i) thermal equilibrium and (ii) lepton number being
small. If these are so then $D(p) \simeq y_{\alpha} G_F^2 T^5 p/2\langle p \rangle_0$. The
fact that $D$ is most of the time just a given external function of temperature and momentum
rather than a dynamically evolving quantity simplifies the numerical solution to the
equations. Observe that the absence of thermal equilibrium is correlated with sufficiently
weak interaction rates, so $D$ is necessarily small in that case relative to ${\bf V}$ (for
relevant choices of oscillation parameters). If, on the other hand, thermal equilibrium
holds, but the neutrino chemical potentials are nonzero and evolving in time, then the
dependence of $D$ on $\rho$ will only be important when the asymmetries are very large. For
interesting choices of oscillation parameters, this only occurs at temperatures approaching
the BBN epoch when collisions are again becoming insignificant. It turns out that in
practice one may neglect all but the leading neutrino asymmetry independent term $y_{\alpha}
G_F^2 T^5 p/2 \langle p \rangle_0$ in the applications we have in mind, a point we will
discuss more fully later on.

Pauli blocking is neglected in the calculation of $D$, and also the repopulation
function $R$ discussed below.  To incorporate this effect, appropriate factors of $(1 -
f)$, where $f(p)\equiv 2\pi^2 N(p)/p^2$, need to be inserted into the integrals.  To
estimate the size of the
error introduced by neglecting Pauli blocking, we can evaluate $f$ at the peak of the
equilibrium momentum distribution, $p \simeq 2.2T$.  Since $f(2.2T)\simeq 0.1$ we
estimate the error to be at most $10\%$.

We now discuss the {\it repopulation} or {\it refilling function} $R(p)$ which controls the
evolution of $P_0(p)$. Since
\begin{equation}
P_0(p) = \frac{N_{\alpha}(p) + N_s(p)}{N^{\text{eq}}(p)},
\label{P0expression}
\end{equation}
the evolution of $P_0$ is governed by processes that deplete or enhance the abundance
of $\nu_{\alpha}$'s with momentum $p$.
The rate of change of $P_0(p)$ clearly receives no contribution from quantally coherent
$\nu_{\alpha} - \nu_s$ oscillations, because the two flavours simply swap.
It therefore equals
the rate at which $\nu_{\alpha}$'s of momentum $p$ are generated by scattering
processes, minus the rate at
which $\nu_{\alpha}$'s of momentum $p$ are scattered out of that momentum value. Its
general form is
\begin{eqnarray}
R(k) & = & \frac{2\pi}{f^{\text{eq}}(k,0)} \int \frac{d^3k'}{(2\pi)^3}\frac{d^3p'}{(2\pi)^3}
\frac{d^3p}{(2\pi)^3} \delta_E(k+p-k'-p')\times \nonumber \\
&\ & \sum_j
\{V^2[\nu_{\alpha}(k),j(p)|\nu_{\alpha}(k'),j(p')][f_{\nu_{\alpha}}(k')f_j(p')
-f_{\nu_{\alpha}}(k)f_j(p)] \nonumber \\
& + &
V^2[\nu_{\alpha}(k),\overline{\nu}_{\alpha}(p)|j(k'),\overline{j}(p')]
[f_j(k')f_{\overline{j}}(p')
-f_{\nu_{\alpha}}(k)f_{\overline{\nu}_{\alpha}}(p)]\},
\label{generalR}
\end{eqnarray}
where
\begin{equation}
f^{\text{eq}}(p,\mu)=\frac{1}{1+\exp (\frac{p-\mu}{T})},
\end{equation}
and $V(i,j|i',j')$ denotes an
interaction matrix element (defined in the Appendix) for
the process $i+j\rightarrow i'+j'$.

As for the decoherence function $D$, the general expression for $R$ in
Eq.(\ref{generalR}) simplifies when
thermal equilibrium holds. For temperatures between neutrino decoupling and 100 MeV,
it is given by
\begin{equation}
R(p) = \Gamma_{\alpha}(p)\left\{
\frac{N^{\text{eq}}(p,\mu_{\alpha})}{N^{\text{eq}}(p,0)}
- \frac{1}{2} P_0(p) [1 + P_z(p)] \right\},
\label{explicitR}
\end{equation}
where $N^{\text{eq}}(p,\mu)$ is the Fermi-Dirac momentum distribution function with
chemical potential $\mu$, and $\mu_{\alpha}$ is the
chemical potential for $\nu_{\alpha}$. This expression follows from Eq.(\ref{generalR})
when all number densities except for $\nu_{\alpha}$ are thermal, and $\nu_{\alpha}$ is
approximately thermal (see the Appendix for the derivation).
The chemical potential $\mu_{\alpha}$
is obtained through the lepton asymmetry using
\begin{equation}
L_{\alpha} \simeq  \frac{T^3}{6n_{\gamma}}\left( \frac{\mu_{\alpha}}{T} \right).
\end{equation}
The physical interpretation of this expression for $R$ is that all weak
interaction processes involving $\nu_{\alpha}$ are tending to send its number density
towards the equilibrium configuration. Equation (\ref{explicitR}) was adopted in
previous works on a heuristic basis \cite{bell}. In fact, it is an exact result under
the
conditions stated above (see the Appendix). Observe that above the neutrino decoupling
temperature, $\nu_{\alpha}$ is approximately in thermal equilibrium and so $R$ is very
small (this issue is discussed more carefully in the next Subsection). Below the neutrino
decoupling temperature, the form given in Eq.(\ref{generalR}) should in principle be
used. For antineutrinos, $\overline{R}$ is obtained from $R$ by replacing $\mu_{\alpha}$ by
$-\mu_{\alpha}$ (when thermal equilibrium holds).

Clearly, the dependence of the neutrino momentum distributions on the repopulation
function will be most sensitive at low temperatures and when the asymmetry has evolved
to large values.  At high temperatures, the weak interaction rates are fast enough
to thermalise the distributions rapidly, spreading out an asymmetry across the
distribution (as described by the chemical potentials).  At lower temperatures, this does
not occur
as efficiently, so the correct form for the repopulation function becomes more
complicated [see Eq.(\ref{generalR})] if
we want to track the actual form of the neutrino momentum distributions. We will discuss the
low temperature ($T \simeq \text{a few MeV}$) regime later on in Subsection
\ref{lowtemp}.

\subsection{Evolution of neutrino asymmetry}

We now focus on extracting the evolution of lepton number or neutrino asymmetry
$L_{\alpha}$ from the Quantum Kinetic Equations. Recall that this is a crucial
quantity because the Wolfenstein $a(p)$ term in the effective matter
potential [see Eq.(\ref{Valpha})] is proportional to a linear combination of
fermion, and in particular neutrino, asymmetries. If this term is appropriately
large, then active-sterile oscillations will be suppressed at low temperatures. Also
recall that $L_e$ affects primordial light element abundances through weak
interaction BBN reaction rates.

In principle, the QKEs (\ref{vecPeqn}) and (\ref{P0eqn})
(plus the antineutrino equations) are solved to indirectly yield
the evolution of the asymmetry defined by Eq.(\ref{asymdef}). It is, however,
very important to obtain simpler approximate evolution equations for $L_{\alpha}$. There
are two reasons for this. First,
it is difficult to understand the qualitative behaviour of the evolution of the
asymmetry from the in-principle QKE procedure {\it per se}. Analytic insight requires
that the evolution of $L_{\alpha}$ be directly considered and approximations
introduced. Second, the direct numerical solution of the full QKEs is computationally
demanding because of the neutrino momentum spread and the MSW resonance phenomenon.

We begin our study of approximate evolution equations by observing that
Eqs.(\ref{relnums}), (\ref{asymdef}) and (\ref{Nn}) combine to produce
\begin{eqnarray}
\frac{d L_{\alpha}}{dt} & = & \frac{1}{2}
\int  \left[ \frac{\partial P_0}{\partial t} (1 + P_z) + P_0 \frac{\partial
P_z}{\partial t} - \frac{\partial
\overline{P}_0}{\partial t}(1 + \overline{P}_z) +
\overline{P}_0 \frac{\partial \overline{P}_z}{\partial t} \right]
\frac{N^{\text{eq}}(p,0) dp}{n_{\gamma}} \nonumber \\
& = & \frac{1}{2}
\int  \left[ 2 \frac{\partial P_0}{\partial t}
- 2 \frac{\partial \overline{P}_0}{\partial t} +
\beta (P_0 P_y - \overline{P}_0 \overline{P}_y) \right]
\frac{N^{\text{eq}}(p,0) dp}{n_{\gamma}} \nonumber \\
& = & \frac{1}{2}
\int \beta (P_0 P_y - \overline{P}_0 \overline{P}_y)
\frac{N^{\text{eq}}(p,0) dp}{n_{\gamma}}
\label{Lev1}
\end{eqnarray}
as the time evolution equation for the $\alpha$-like neutrino asymmetry
$L_{\alpha}$. We have used the fact that $N^{\text{eq}}(p,0)dp/n_{\gamma}$ is
time independent, and also conservation of lepton number through
\begin{equation}
\int (P_0 - \overline{P}_0) \frac{N^{\text{eq}}(p,0) dp}{n_{\gamma}}
= \frac{n_{\alpha} + n_s - \overline{n}_{\alpha} - \overline{n}_s}{n_{\gamma}} = \text{constant}.
\end{equation}
So far, no approximations have been made (except
that possible tiny flavour changing neutral current effects for massive
neutrinos have been neglected.\footnote{These effects will always be small, with
their exact size being model dependent.})
Using Eq.(\ref{P0expression}) and the corresponding expression for antineutrinos,
Eq.(\ref{Lev1}) implies that
\begin{equation}
\frac{dL_{\alpha}}{dt} = \frac{1}{2n_{\gamma}} \int \beta\left[(N_{\alpha} + N_s) P_y -
(\overline{N}_{\alpha} + \overline{N}_s) \overline{P}_y\right] dp.
\label{Lev2}
\end{equation}
To proceed, we need approximate solutions for $P_y$ and $\overline{P}_y$.
These are obtained by solving the QKEs for ${\bf P}$ and the corresponding
antineutrino equations.
In expanded form, the QKEs (\ref{vecPeqn}) are
\begin{eqnarray}
\frac{\partial P_x}{\partial t} & = & - \lambda P_y - D P_x - \frac{P_x}{P_0}
\frac{\partial P_0}{\partial t},\nonumber\\
\frac{\partial P_y}{\partial t} & = & \lambda P_x - \beta P_z - D P_y -
\frac{P_y}{P_0}\frac{\partial P_0}{\partial t},\nonumber\\
\frac{\partial P_z}{\partial t} & = & \beta P_y + \frac{1 - P_z}{P_0} \frac{\partial
P_0}{\partial t},
\label{xyz}
\end{eqnarray}
which also require Eq.(\ref{P0eqn}).

The qualitative character of the evolution goes through distinct phases. We first
discuss the high temperature initial conditions. As stated previously, it is expected
that the initial abundance of sterile species is very close to zero. This implies
that $P_z \simeq 1$ at high $T$, so that $N_s \simeq 0$. At temperatures above the
neutrino decoupling temperature, the neutrinos are by definition in thermal
equilibrium, so the repopulation function R equals zero according to
Eq.(\ref{explicitR}). This means that $P_0$ is a constant which depends on the
initial neutrino chemical potential and hence on the initial neutrino asymmetry. We
will also take the initial neutrino asymmetry to be negligible (it could be of the
order of the baryon asymmetry, for instance\footnote{A natural possibility is that
the primordial neutrino asymmetry is produced by a similar mechanism to the baryon
asymmetry. It is of course also possible that some relatively high temperature
mechanism operates that produces a primordial neutrino asymmetry that is much larger
than the baryon asymmetry. However, we will not consider this scenario in this
paper.}). If this is so, then $P_0 \simeq 1$.

The first phase occurs at high temperatures. With the asymmetry set to its initial value
of zero, the function $\lambda$ is dominated by the finite temperature term of the
effective potential
\begin{equation}
V_{\alpha} \simeq \frac{\Delta m^2}{2p} b(p).
\end{equation}
Replacing $p$ by its thermal average, we see that the magnitude of this term increases
with temperature
as $T^5$. The function $\beta$ effectively goes to zero at sufficiently high temperature
because it scales as $T^{-1}$. The decoherence function $D$, like $\lambda$, scales as
$T^5$. (Note that, numerically, $\left|V_{\alpha}/D\right| \simeq 60\ \text{or}\ 24$
for $\alpha = e$ or $\mu/\tau$, respectively.) Setting $P_0$ to be constant, and
neglecting
$\beta$, the QKEs (\ref{xyz}) become
\begin{equation}
\frac{\partial P_x}{\partial t} \simeq - \lambda P_y - D P_x,\quad
\frac{\partial P_y}{\partial t} \simeq \lambda P_x - D P_y,\quad
\frac{\partial P_z}{\partial t} \simeq 0.
\label{xyz-highT}
\end{equation}
Therefore $P_z(t)$ is frozen at its initial value $P_z(t) \simeq 1$, and no sterile
neutrinos are produced. Writing
\begin{equation}
P_x + i P_y \equiv |P|e^{i\phi},
\end{equation}
the approximate QKEs (\ref{xyz-highT}) become
\begin{equation}
\frac{\partial |P|}{\partial t} \simeq -D |P|,\quad \frac{\partial \phi}{\partial t}
\simeq \lambda,
\end{equation}
which are trivially solved to yield
\begin{equation}
|P(t)|e^{i\phi(t)} \simeq |P(0)| e^{i\phi(0)} e^{-\int^t_0 D(t') dt'}
e^{i\int^t_0 \lambda(t') dt'}.
\end{equation}
Since, at high temperatures, $D$ is large compared to the expansion rate of the universe
we see that $|P|$ and hence also $P_x$ and $P_y$ get exponentially damped to zero very
rapidly from any initial values. (Observe that they rapidly oscillate about zero as they do
so due to the large $\lambda$.) The system is therefore completely incoherent
(``in a mixed state'') at high temperatures, and frozen (Quantum Zeno or Turing Effect).

The third phase occurs at very low temperatures where the collisional terms $D$ and $b$
are negligible. The neutrino subsystem then evolves as a coherently oscillating
subsystem which is also coupled through $R$ to the background. The oscillations are
matter-affected if a sufficiently large neutrino asymmetry $L^{(\alpha)}$ has been
created by this time.  This is the regime relevant for the BBN epoch, for interesting
choices of oscillation parameters. We will discuss this regime in a later Subsection.

Intermediate between these two limiting regimes lies the second phase. During this
phase, the system emerges from its frozen initial state. The evolution is driven by an
interplay between collisions, which are still rapid compared to the expansion rate, and
oscillations, which begin to affect the neutrino number densities through a
non-negligible $\beta$ (recall that $\beta$ is proportional to $\sin^2 2 \theta_0$ where
$\theta_0$ is the vacuum mixing angle). This is the regime of the ``static
approximation'' introduced in Ref.\cite{longpaper}. It is during this epoch that
neutrino asymmetries will evolve to values orders of magnitude higher than the baryon
asymmetry provided the oscillation parameters are in the appropriate region.  We will
now provide new analytical insight into this phenomenon by deriving in a new way the
very useful approximate form for the asymmetry evolution equation considered in
Ref.\cite{longpaper}. We will see that the static approximation is actually a combination
of an adiabatic approximation for {\it partially incoherent} oscillations, and a small
$\beta$ expansion.
This point was not crisply realised hitherto.

We now take the first crucial step, that
\begin{equation}
\frac{\partial P_0}{\partial t} \simeq 0
\label{P0const}
\end{equation}
continues to be a good approximation.
With this approximation, the QKEs (\ref{xyz}) simplify to the homogeneous equations
\begin{equation}
\frac{\partial}{\partial t}
\left( \begin{array}{c}
P_x \\ P_y \\ P_z
\end{array} \right) \simeq
\left( \begin{array}{ccc}
-D\ & -\lambda\ & 0 \\ \lambda\ & -D\ & -\beta \\ 0\ & \beta\ & 0
\end{array} \right)
\left( \begin{array}{c} P_x \\ P_y \\ P_z \end{array} \right),
\label{approxeq}
\end{equation}
or, in a self-evident matrix notation,
\begin{equation}
\frac{\partial{\bf P}}{\partial t} \simeq {\cal K} {\bf P}.
\end{equation}
The continued justification for Eq.(\ref{P0const}) is as follows: Above the neutrino decoupling
temperature, most of the $\nu_{\alpha}$ ensemble is in thermal equilibrium. From
Eq.(\ref{explicitR}) we see that $R$ is therefore zero or very small. One might be tempted
to conclude that in fact $R$ is exactly zero above the neutrino decoupling temperature.
This, however, cannot strictly be the case.
Suppose $\nu_{\alpha}$'s of some given resonance momentum $p_{\text{res}}$ oscillate strongly
to $\nu_s$'s.
Instantaneously, $N_{\alpha}(p_{\text{res}})$ goes to zero or close to it, but
$P_0(p_{\text{res}})$ does not change due to
these oscillations because of the generation of a nonzero $N_s(p_{\text{res}})$. However, the
absent momentum mode of $\nu_{\alpha}$'s is quickly repopulated from the background medium,
leading clearly to an overall nonzero value for $\frac{\partial P_0}{\partial t}$ evaluated
at the momentum $p_{\text{res}}$. We would therefore expect that $R$ is very small above
the neutrino decoupling
temperature, except when the oscillations are very strong. This means that the approximation
in Eq.(\ref{P0const}) will be a good one except possibly in the centre of an MSW resonance
when oscillations might be rapid. The word ``might'' is
used because the oscillations will be non-adiabatic in some regions of parameter space. So,
we proceed
with the understanding that our approximation scheme might not be strictly valid in the
centre of a resonance. We will have cause to revisit the centre of the resonance later
on.

To solve Eq.(\ref{approxeq}), we first introduce the ``instantaneous diagonal basis''
through
\begin{equation}
\left( \begin{array}{c}
Q_1 \\ Q_2 \\ Q_3
\end{array} \right) \equiv {\bf Q} = {\cal U} {\bf P},
\end{equation}
where ${\cal U}$ is a time-dependent matrix that diagonalises ${\cal K}$,
\begin{equation}
\label{kd}
{\cal K}_d \equiv \text{diag}(k_1, k_2, k_3) = {\cal U} {\cal K} {\cal U}^{-1},
\end{equation}
with $k_{1,2,3}$ being eigenvalues.
In the instantaneous diagonal basis, Eq.(\ref{approxeq}) becomes
\begin{equation}
\frac{\partial {\bf Q}}{\partial t} \simeq {\cal K}_d {\bf Q} - {\cal U} \frac{\partial
{\cal U}}{\partial t}^{-1} {\bf Q}.
\label{udu}
\end{equation}
This equation resembles, but is not the same as, the MSW evolution equation written
in the instantaneous mass eigenstate basis with a varying background density. In that
case, the adiabatic approximation sees the time derivative of the instantaneous
mixing matrix set to zero.

Our next important approximation after Eq.(\ref{P0const}) is to analogously set
\begin{equation}
\frac{\partial {\cal U}}{\partial t}^{-1} \simeq 0,
\end{equation}
so that the evolution equation becomes
\begin{equation}
\frac{\partial {\bf Q}}{\partial t} \simeq {\cal K}_d {\bf Q}.
\label{dada}
\end{equation}
This is clearly within the family of adiabatic-like approximations. (We defer the
discussion of the region of applicability of this approximation to a later subsection.) The
difference with
the usual adiabatic approximation for MSW evolution is the existence of quantum decoherence
through a nonzero $D$. We will now show that Eq.(\ref{dada}) reproduces the approximate
evolution equation derived in Ref.\cite{longpaper} using what was termed the static
approximation. {\it
The static approximation of Ref.\cite{longpaper} is therefore revealed to be just the
adiabatic
approximation for a partially incoherent system of neutrinos undergoing matter affected
oscillations (in the small $\beta$ limit --- see below).} (Note that this
type of idea was briefly discussed in Ref.\cite{barbieri}.)

Equation (\ref{dada}) can be formally solved to yield
\begin{equation}
\left[ \begin{array}{c}
Q_1(t) \\ Q_2(t) \\ Q_3(t)
\end{array} \right] =
\left( \begin{array}{ccc}
e^{\int^t_0 k_1(t')dt'}\ & 0\ & 0 \\
0\ & e^{\int^t_0 k_2(t')dt'}\ & 0 \\
0\ & 0\ & e^{\int^t_0 k_3(t')dt'}
\end{array} \right)
\left[ \begin{array}{c}
Q_1(0) \\ Q_2(0) \\ Q_3(0)
\end{array} \right],
\end{equation}
which in turn implies that
\begin{equation}
\left[ \begin{array}{c}
P_x(t) \\ P_y(t) \\ P_z(t)
\end{array} \right] = {\cal U}^{-1}(t)
\left( \begin{array}{ccc}
e^{\int^t_0 k_1(t')dt'}\ & 0\ & 0 \\
0\ & e^{\int^t_0 k_2(t')dt'}\ & 0 \\
0\ & 0\ & e^{\int^t_0 k_3(t')dt'}
\end{array} \right) {\cal U}(0)
\left[ \begin{array}{c}
P_x(0) \\ P_y(0) \\ P_z(0)
\end{array} \right].
\label{staticsoln}
\end{equation}
The mixing matrix ${\cal U}^{-1}$ is obtained
by placing the normalised eigenvectors $\kappa_i$ in the columns, where
\begin{equation}
\kappa_i = \frac{1}{{\cal N}_i}
\left[ \begin{array}{ccc}
1\ \\ -\frac{D + k_i}{\lambda}\ \\ -\beta\frac{D + k_i}{\lambda k_i}
\end{array} \right]
\label{evectors}
\end{equation}
and
\begin{equation}
\label{normalisation}
{\cal N}_i = \frac{1}{\lambda |k_i|}\sqrt{\lambda^2 |k_i|^2 + (\beta^2
        + |k_i|^2) |D + k_i|^2}.
\end{equation}
The inverse matrix ${\cal U}$ is then composed of the row vectors $v_i$, where
\begin{equation}
v_i = - {\cal N}_i k_i \left( \begin{array}{ccc}
    \frac{1}{D} \prod\limits_{j \neq i} \frac{D + k_j}{k_i - k_j},\
    & \lambda \prod\limits_{j \neq i} \frac{1}{k_i - k_j},\
    & \frac{\lambda}{\beta D} \prod\limits_{j \neq i}
        \frac{k_j}{k_i - k_j}
    \end{array} \right).
\end{equation}
One of the eigenvalues will be real and the other two will be a complex
conjugate pair: $k_2 = k_1^*$ and $k_3 = k_3^*$.

In order to progress further, we need to calculate the eigenvalues. Since we are dealing with a $3 \times 3$ matrix, this is somewhat awkward algebraically.
Fortunately, the
small $\beta$ limit is of great interest. The numerical details of the small $\beta$
limit will be given in a later
subsection. In this limit it is easy to show that the eigenvalues are approximately given by
\begin{equation}
k_1 = k_2^* = -D + i\lambda + O(\beta^2),\qquad k_3 = -\frac{\beta^2 D}{D^2 +
\lambda^2} + O(\beta^4).
\label{evalues}
\end{equation}
Since we are still in the regime where the decoherence parameter $D$ is significant,
it follows that
\begin{equation}
e^{\int_0^t k_{1,2}(t') dt'} \simeq 0.
\end{equation}
Furthermore,
\begin{equation}
e^{\int_0^t k_3(t') dt'} = 1 + O(\beta^2).
\end{equation}
Combining these results with Eqs.(\ref{staticsoln}-\ref{evalues}) we
obtain that
\begin{equation}
P_y(t) \simeq - \frac{\beta D}{D^2 + \lambda^2} P_z(t) + O(\beta^3).
\end{equation}
For the antiparticle system we obtain a similar expression, so the
neutrino asymmetry evolution equation (\ref{Lev2}) becomes
\begin{equation}
\frac{d L_{\alpha}}{dt} \simeq \frac{1}{2n_{\gamma}} \int \beta^2 \left[
\frac{\overline{D}(\overline{N}_{\alpha} - \overline{N}_s)}{\overline{D}^2 +
\overline{\lambda}^2} -
\frac{D(N_{\alpha} - N_s)}{D^2 + \lambda^2}
\right] dp,
\label{Lev3}
\end{equation}
having used
\begin{equation}
P_z = \frac{N_{\alpha} - N_s}{N_{\alpha} + N_s}
\end{equation}
and the corresponding antineutrino expression. From Eq.(\ref{Lev3}) we see that lepton
number is constant if the number density distributions of active and sterile species are
equal. This is to be expected, because oscillations effectively do nothing in that case.

Since we are working within the $R = \overline{R} = 0$ approximation, we now substitute
$N^{\text{eq}}(p,\mu)$ for $N_{\alpha}$ and $N^{\text{eq}}(p,-\mu)$ for
$\overline{N}_{\alpha}$. (The chemical potentials of $\nu_{\alpha}$ and
$\overline{\nu}_{\alpha}$ are equal and opposite because of the rapid $\nu_{\alpha}
\overline{\nu}_{\alpha} \leftrightarrow e^{+}e^{-}$ process.) Since we are taking the
initial lepton number to be small, we will also expand $N^{\text{eq}}(p,\mu)$ to first order
in
\begin{equation}
\frac{\mu}{T} \simeq \frac{6 n_{\gamma} L_{\alpha}}{T^3},
\end{equation}
to obtain
\begin{equation}
N^{\text{eq}}(p,\mu) \simeq N^{\text{eq}}(p,0) + N^{\text{eq}}(p,0) \frac{e^{p/T}}{1 +
e^{p/T}}\frac{\mu}{T}.
\end{equation}
This allows us to form the quantities $N^+_{\alpha}$ and $N^-_{\alpha}$, which
are given by
\begin{eqnarray}
N^+_{\alpha}=\frac{1}{2}(N_{\alpha}+\overline{N}_{\alpha})&=&
N^{\text{eq}}(p,0) + O(L_{\alpha}^2), \nonumber \\
N^-_{\alpha}=\frac{1}{2}(N_{\alpha}-\overline{N}_{\alpha})&=&
L_{\alpha}\frac{12\zeta(3)}{\pi^2}\frac{e^{p/T}}{1+e^{p/T}} N^{\text{eq}}(p,0) +
O(L_{\alpha}^3).
\end{eqnarray}
The asymmetry evolution equation (\ref{Lev3}) becomes,
\begin{equation}
\label{Lev4}
\frac{d L_{\alpha}}{dt} \simeq \frac{\pi^2}{2\zeta(3)T^3} \int
\frac{s^2 \Gamma_{0}
a(c-b)}{[x+(c-b+a)^2][\overline{x}+(c-b-a)^2]}(N^+_{\alpha}-N^+_s)dp
+\Delta +\delta,
\end{equation}
where $c \equiv \cos 2\theta_0$, $s \equiv \sin 2\theta_0$,
\begin{equation}
x \equiv \left[ \frac{p \Gamma_{\alpha}(p)}{\Delta m^2} \right]^2,\qquad
\overline{x} \equiv \left[ \frac{p \overline{\Gamma}_{\alpha}(p)}{\Delta m^2} \right]^2,
\label{xdefn}
\end{equation}
and
\begin{equation}
\Gamma_{0} \equiv \langle \Gamma_{\alpha} \rangle_0 \frac{p}{\langle p \rangle_0}
\end{equation}
is the collision rate with the chemical potential set to zero.
The quantity $\Delta$ is an $O(L_{\alpha})$ correction term given by
\begin{equation}
\Delta = - \frac{\pi^2}{4\zeta(3)T^3} \int
\frac{s^2\Gamma_{0} [x_0 + a^2 + (b-c)^2]}
{[x + (c-b+a)^2][\overline{x} + (c-b-a)^2]}(N^-_{\alpha}-N^-_s)dp,
\end{equation}
and $\delta$ is an additional $O(L_{\alpha})$ correction term which arises
from
allowing $D$ to be different from $\overline{D}$,
\begin{equation}
\delta \simeq -z_{\alpha}L_{\alpha}\frac{\pi^2}{4\zeta(3)T^3} \int
\frac{s^2\Gamma_0[-x_0+a^2+(b-c)^2]}
{[x+(c-b+a)^2][\overline{x}+(c-b-a)^2]}(N^+_{\alpha}-N^+_s)dp,
\end{equation}
where $x_0$ is $x$ as in Eq.(\ref{xdefn}) but with $\Gamma_{0}$ in place of
$\Gamma_{\alpha}$.
The evolution equation (\ref{Lev4}), including the correction term $\Delta$ but excluding
$\delta$, was
studied in detail in Ref.\cite{longpaper}. From Eq.(\ref{Lev4}) we see that the effect
of having
different collision rates for the neutrinos and antineutrinos is to introduce the
additional correction, $\delta$, as well as a small correction through $x$ and
$\overline{x}$ to the denominators.

The $\Delta$ and $\delta$ correction terms, as well as the $O(L_{\alpha})$ corrections
to $x$ and $\overline{x}$ in the denominators may readily be incorporated into the numerical
study
of the evolution equation (\ref{Lev4}), however to a good approximation we may neglect
these corrections.
It turns out that $\delta$ only amounts to a small contribution such that
\begin{equation}
\frac{|\delta|}{|\Delta|} \sim 0.1.
\end{equation}
The corrections to the denominators are very small, even at a resonance.  At the
initial MSW resonance where $a \simeq 0$, $(b-c) \simeq0$, $[x + (c-b+a)^2][\overline{x} +
(c-b-a)^2]\rightarrow x_0^2 + O(L_{\alpha}^2)$.

The approximate evolution equation (\ref{Lev4}) has proven to be very useful for gaining
analytical insight and indeed for performing numerical work. The $\delta = 0$ version was
studied in depth in Ref.\cite{longpaper}. The derivation presented above shows for the first
time that the full glory of Eq.(\ref{Lev4}) follows from first principles (that is, the
QKEs). In Ref.\cite{longpaper}, a heuristic derivation based on a Pauli-Boltzmann--like
approximation was used.  A minor issue associated with the definition of
$x$ arises on closer examination.  In Ref.\cite{longpaper}, this
quantity contains an additional term $\sin^2 2 \theta_0$ that is absent in
the present definition.  It turns out that keeping higher order terms
in $\beta$ in the derivation of
Eq.(\ref{Lev3}) does not in fact reproduce this missing term as
one might hope.  The mystery of the missing $\sin^2 2 \theta$ seems to require
an explanation at a deeper level, as yet unprobed.
Note also that when the sterile neutrino number density is negligible,
Eq.(\ref{Lev4}) becomes a self-contained non-linear differential equation describing the
evolution of the single momentum-independent variable $L_{\alpha}$.  It is a considerable
simplification over the full QKEs which are coupled differential equations for the eight
momentum-dependent functions ${\bf P}$, $P_0$, $\overline{{\bf P}}$ and $\overline{P}_0$.

In order to integrate Eq.(\ref{Lev4}) when the sterile neutrino number density is not
negligible, an evolution equation for $N_s$ is needed. Using
\begin{equation}
\frac{N_s(p)}{N^{\text{eq}}(p,0)} = \frac{1}{2} P_0(p) [1 - P_z(p)]
\end{equation}
and the QKEs with the same approximations as above, we easily obtain
\begin{eqnarray}
\frac{d}{dt}\left(\frac{N_s(p)}{N^{\text{eq}}(p,0)}\right) & \simeq &
-\frac{1}{2} \beta P_0(p) P_y(p) \nonumber \\
& \simeq & \frac{1}{4}\left(\frac{N_{\alpha}(p)-N_s(p)}{N^{\text{eq}}(p,0)}\right)
\frac{\Gamma_{\alpha} s^2}{x+(c-b+a)^2} \nonumber \\
&\simeq&  \frac{1}{4} \frac{\Gamma_{0} s^2}{x+(c-b+a)^2} \nonumber \\
&\times& \left[1 -\frac{N_s(p)}{N^{\text{eq}}(p,0)}
+ L_{\alpha} \left[\frac{6 n_{\gamma}}{T^3} \frac{e^{p/T}}{1 +
e^{p/T}} - z_{\alpha}\left(1 -\frac{N_s(p)}{N^{\text{eq}}(p,0)}\right) \right]+  O(L_{\alpha}^2) \right],
\label{sterev}
\end{eqnarray}
as the required evolution equation. The sterile antineutrino equation obviously has a
similar form with the substitutions $a \to -a$, $L_{\alpha} \to -L_{\alpha}$ and $N_s \to
\overline{N}_s$. Equations (\ref{Lev4}), (\ref{sterev}) plus the sterile antineutrino
equation are a useful set of coupled equations that approximately describe the evolution of
lepton number.

\subsection{Qualitative features of lepton number evolution}

For completeness, we now briefly recall the qualitative features of
lepton number evolution which are revealed by Eq.(\ref{Lev4}). For
further discussion see Ref.\cite{longpaper}.
Initially, neutrino
asymmetries are small and hence the term $a$ is small. A critical factor in the
subsequent evolution of $L_{\alpha}$ is the overall sign of its derivative. (Note also
the distinction between $L_{\alpha}$ and $L^{(\alpha)}$.)
If it is of the opposite sign to $L^{(\alpha)}$, then the asymmetry
$L_{\alpha}$ will evolve such that
\begin{equation}
L^{(\alpha)} \to 0.
\end{equation}
If, on the other hand, the derivative is of the same sign as $L^{(\alpha)}$,
then the evolution of $L_{\alpha}$ will be such as to increase
$L^{(\alpha)}$. For $\Delta m^2 > 0$, the former situation always obtains. No
explosive creation of lepton number can occur. However, when $\Delta m^2 < 0$ the
situation is more complicated. (Observe that for small mixing, $\Delta m^2 < 0$
means
loosely that the sterile neutrino is less massive than the active neutrino). In this
case the quantity $a$ has the same sign as $L^{(\alpha)}$.
The quantity $b$ is always positive, but it is a decreasing function of time.
The sign of the derivative is therefore controlled by $(c-b)$. At a given temperature
$T$, the function $(c-b)$ equals zero for neutrinos and antineutrinos of momentum
$p_c$, where
\begin{equation}
p_c = \frac{\pi M_W}{2 T^2}\sqrt{-\frac{\Delta m^2
c}{\sqrt{2}\zeta(3)A_{\alpha}G_F}}.
\label{pc}
\end{equation}
Neutrinos and antineutrinos with momenta less than $p_c$ make a positive contribution
to the righthand side of Eq.(\ref{Lev4}), whereas those with momenta greater
than $p_c$ make a negative contribution. The critical momentum $p_c$ is very
small at high $T$ but increases with decreasing $T$, which means that the derivative
changes sign from negative to positive once $T$ is small enough for $p_c$ to be near
the peak of the Fermi-Dirac distribution. (For
maximal mixing, $c = 0$ and the derivative is always negative. The creation of lepton
number occurs due to {\it small angle} oscillations.) Since $a$ is
initially very small, the equation
\begin{equation}
c - b = 0
\end{equation}
is also the MSW resonance condition for both neutrinos and antineutrinos. The
denominator in the integrand of Eq.(\ref{Lev4}) is thus at a minimum for
neutrinos and antineutrinos with $p = p_c \simeq p_{\text{res}}$. The resonance
momentum $p_{\text{res}}$ is generally found by solving
\begin{equation}
c - b \pm a = 0,
\end{equation}
where the plus (minus) sign pertains to neutrinos (antineutrinos). Provided the
vacuum mixing angle is small enough, the critical
temperature $T_c$ at which the sign of the derivative changes is well approximated by
taking $p_c \simeq 2.2 T_c$ which corresponds to the peak of the Fermi-Dirac
distribution. From Eq.(\ref{pc}), it is given approximately by
\begin{equation}
T_c \simeq \left[ \frac{\pi M_W}{4.4}\sqrt{-\frac{\Delta m^2
c}{\sqrt{2}\zeta(3)A_{\alpha}G_F}} \right]^{1/3} \simeq 15(18) \left(\cos 2\theta_0
\frac{|\Delta m^2|}{\text{eV}^2}\right)^{1/6}\ \text{MeV},
\label{Tc}
\end{equation}
for $\nu_e - \nu _s$ ($\nu_{\mu,\tau} - \nu _s$) oscillations. (Numerically, the
critical temperature is found to be slightly higher than this estimate \cite{longpaper}.
Also, when the vacuum mixing angle is sufficiently large, a non-negligible sterile
neutrino
number density is produced, causing the actual $T_c$ to differ from this estimate.)

Consider now the parameter space region
\begin{equation}
\cos\theta \simeq 1,\quad |\Delta m^2| > 10^{-4}\ \text{eV}^2,
\end{equation}
chosen so that the mixing angle is small and so that
$T_c$ occurs above the neutrino decoupling temperature.
At $T \simeq T_c$, explosive exponential growth of lepton number begins\footnote{Note
that while the mixing angle $\theta$ should be small, it must be large enough so that
$\sin^2 2\theta \stackrel{>}{\sim} 5 \times 10^{-10} (\text{eV}^2/|\Delta m^2|)^{1/6}$
is satisfied, otherwise the partially incoherent oscillations are not strong enough to
create lepton number. See Ref.\cite{longpaper} for more details.}
because Eq.(\ref{Lev4}) is then of the form
\begin{equation}
\frac{dL_{\nu_{\alpha}}}{dt}\  \propto\  + L_{\nu_{\alpha}},
\end{equation}
with the proportionality factor augmented by the MSW resonances in the denominator of
the integrand. After a short time, lepton number has grown sufficiently for the
quantity $a$ to no longer be negligible. Non-linearity through the
denominator then alters the qualitative character of the evolution from exponential
growth into a slower phase governed approximately by
\begin{equation}
a(\langle p \rangle) \simeq \pm \cos 2\theta_0,
\end{equation}
depending on whether the sign of the asymmetry is positive or negative. This equation
is the resonance condition for neutrinos (antineutrinos) when $|a| \gg |b|$.

As the temperature continues to decrease, the collision dominated epoch begins to give
way to an epoch in which coherent oscillations take over as the dominant driver of
neutrino evolution. By the time the neutrino decoupling temperature is reached, coherent
oscillations certainly dominate over collisions.  This is the regime relevant for the
BBN epoch.  Equation (\ref{Lev4}) no longer provides a good description of lepton number
evolution.  The evolution of lepton number is, of course, in principle obtained by
solving Eqs.(\ref{vecPeqn}) and (\ref{P0eqn}), using the general expression given in
Eq.(\ref{generalR}) for the function $R$.  (Numerically, setting $D, b \simeq 0$ is a
good approximation in this regime.)  This in-principle procedure has, to our knowledge,
not yet been attempted in practice.  Instead, the low temperature calculations that have
actually been performed have utilised the usual adiabatic approximation for fully
coherent matter-affected oscillations, and the growth of the asymmetry has been
calculated using an equation which determines how fast the resonance momentum moves
through the distribution converting active into sterile neutrinos. The derivation of
this equation from the QKEs will be presented, for the first time, in Subsection
\ref{lowtemp}. Furthermore, the
high-temperature expression for $R$ as given in Eq.(\ref{approxR}) has been adopted as
an approximation to the much more complicated and rigorous expression defined in
Eq.(\ref{generalR}). We may estimate the theoretical error that this approximate
repopulation method introduces into the final result by comparing the value of the
asymmetry obtained in the manner described above with that obtained by integrating over
the neutrino momentum distributions to find the total number densities. If everything is
consistent, these should of course agree.  However, for the calculations presented in
Ref.\cite{bell}, we find discrepancies of up to $10$ to
$20\%$.  The repopulation should not change the total asymmetry, only distribute it from
the particular momentum state at which it is created, across the momentum distribution.
The discrepancy, which must be due to inconsistencies in the repopulation process, is a
good indication of the theoretical error in all of the calculations
\cite{Nnufv,bell} performed thus far.
This theoretical error is small enough for the conclusions so far reached to
be essentially unchanged.

\subsection{Region of applicability for the approximations}

The validity of Eq.(\ref{Lev3}) rests on the assumptions that the quantity
$\beta$ is small relative to the size of $D$ and $\lambda$, and that their
rates of change with respect to time are negligible.  We now examine the
regions where these conditions are satisfied.

The small $\beta$ approximation is sufficiently accurate provided that
\begin{equation}
\left| \frac{\beta}{\sqrt{D^2 + \lambda^2}} \right| \ll 1.
\end{equation}
The ``small $\beta$ expansion'' is more properly to be thought of as an expansion in the
dimensionless quantity $\beta/\sqrt{D^2 + \lambda^2}$. In order to explore the bound,
we will replace $p$ by its thermal average $\simeq 3.15 T$.

Away from resonance, $|\lambda| \gg D$, so the small $\beta$ expansion holds provided
that $|\beta| \ll |\lambda|$. For $T > T_c$, $\lambda \simeq \Delta m^2 b/2p$, which
means that the small $\beta$ limit requires
\begin{equation}
\frac{|\Delta m^2|}{\text{eV}^2} \sin 2\theta_0 \ll 2 \times 10^{-7}\ (6 \times 10^{-8})
\left( \frac{T}{\text{MeV}} \right)^6,
\label{smallbeta1}
\end{equation}
for $\alpha = e$ and $\mu/\tau$, respectively.
We would like the small $\beta$ limit to hold just prior to lepton number creation at $T
= T_c$. Putting $T = T_c$, as given by Eq.(\ref{Tc}), in Eq.(\ref{smallbeta1}) we see
that $|\Delta m^2|$ cancels out of the inequality, leaving
\begin{equation}
\tan 2\theta_0 \ll 1
\end{equation}
as the constraint. Note that the use of Eq.(\ref{Tc}) presupposes that the mixing angle
is sufficiently small for the sterile neutrino number density to be small, so this is
not a serious constraint.

In the centre of the lepton-number creating resonance at $T = T_c$, $\lambda$
momentarily goes to zero. The small $\beta$ expansion will be valid in the centre of the
resonance, provided the more stringent constraint $|\beta| \ll D$ holds.
This translates into the requirement that
\begin{equation}
\label{smallbeta}
\left(\frac{|\Delta m^2|}{\text{eV}^2}\right) \sin 2\theta_0
\ll 4 \times  10^{-10} y_{\alpha}
\left( \frac{T_c}{\text{MeV}} \right)^6
\simeq 10^{-9} \left( \frac{T_c}{\text{MeV}} \right)^6.
\end{equation}
Using Eq.(\ref{Tc}), we find that $|\Delta m^2|$ cancels out of the inequality, leaving
the constraint
\begin{equation}
\tan 2\theta_0 \ll 10^{-2}
\end{equation}
on the vacuum mixing angle.

We now discuss the applicability of the adiabatic-like
approximation which assumes the complete negligibility of the term
${\cal U} \frac{\partial {\cal U}}{\partial t}^{-1}$ that appears in Eq.(\ref{udu}).
Evaluated explicitly in the small $\beta$ limit, this hitherto discarded
quantity takes on the form
\begin{equation}
{\cal U} \frac{\partial {\cal U}}{\partial t}^{-1}
= \left( \begin{array}{ccc}
        W & Y^* &  - Z^* \\
        Y & W^* & - Z \\
        Z^* & Z & X \\
        \end{array} \right),
\end{equation}
with
\begin{eqnarray}
W & = & \frac{1}{2 \lambda} \frac{\beta D}{D^2 + \lambda^2}
    \frac{D - i 3 \lambda}{\lambda + i D} \frac{d \beta}{dt} \nonumber \\
    & + &  \frac{i \beta^2}{D^2 + \lambda^2} \left\{ \frac{dD}{dt} \left[
            \frac{1}{4 \lambda}
        + \frac{(\lambda - iD)(D^2 - \lambda^2)}
            {(D^2 + \lambda^2)^2} \right]
    + \frac{d \lambda}{dt}
        \left[ \frac{D}{4 \lambda^2}
                + \frac{2 \lambda D (\lambda - iD)}
    {(D^2 + \lambda^2)^2} \right] \right\}  + O(\beta^3), \nonumber \\
X & = & \frac{-2 \beta D^2}{(D^2 + \lambda^2)^2} \frac{d \beta}{dt}
    + \frac{2 \beta^2 D}{(D^2 + \lambda^2)^3}
    \left[(D^2 - \lambda^2) \frac{dD}{dt}
    + 2 \lambda D \frac{d \lambda}{dt} \right] + O(\beta^3), \nonumber \\
Y & = & \frac{i}{2 \lambda} \frac{\beta D}{D^2 + \lambda^2} \frac{d \beta}{dt}
    - \frac{i}{4 \lambda^2} \frac{\beta^2}{D^2 + \lambda^2}
        \left( \lambda \frac{dD}{dt} + D \frac{d \lambda}{dt} \right)
        + O(\beta^3), \nonumber \\
Z & = & \frac{1}{\sqrt{2}} \frac{1}{(\lambda - i D)} \frac{d \beta}{dt}
    - \frac{1}{\sqrt{2}} \frac{\beta}{(\lambda - i D)^2}
    \left( - i \frac{dD}{dt} + \frac{d \lambda}{dt} \right) + O(\beta^2),
\end{eqnarray}
to the lowest order in $\beta$.  These non-vanishing diagonal and
off-diagonal entries contribute to corrections to both the eigenvalues
and eigenvectors of the matrix ${\cal K}_d$ in Eq.(\ref{kd}).
Thus our immediate task is to
ensure that these corrections are small in the region of interest.

The $33$ entry of the matrix ${\cal U} \partial {\cal U}^{-1}/\partial t$ represents a
``first order'' correction to the eigenvalue $k_3$.  Imposing the
condition
\begin{equation}
\left| \frac{ \left( {\cal U} \frac{\partial {\cal U}}{\partial t}^{-1} \right)_{33}}
        {k_3} \right| \ll 1,
\end{equation}
and rendering it into a more illuminating form,
\begin{equation}
\label{whocares}
(D^2 + \lambda^2)^2 \gg \left| 2 D
\frac{1}{\beta} \frac{d \beta}{dt} (D^2 + \lambda^2)
    - 2 \left[ (D^2 - \lambda^2) \frac{dD}{dt} +
        2 \lambda D \frac{d \lambda}{dt} \right] \right|,
\end{equation}
we find that
\begin{equation}
\label{Tbound} T \gg \frac{4.3}{y_{\alpha}^{1/3}} \, \text{MeV}
\simeq 3\, \text{MeV},
\end{equation}
where the momentum $p$ has been replaced with the mean momentum
$\langle p \rangle$.  Note that
in deriving the above, we have set $\lambda \simeq 0$, i.e., the
resonance condition, and we have used the relation
$t \simeq m_{Planck}/11T^2$.  Thus Eq.(\ref{Tbound}) purports the failure
of the static approximation for resonances occurring at temperatures below
roughly $3$ MeV.  Together with the resonance condition $c \simeq b \simeq 1$,
Eq.(\ref{Tbound}) effectively places a lower limit on $|\Delta m^2|$, that is
\begin{equation}
\frac{|\Delta m^2|}{\text{eV}^2} > 8.7 \times 10^{-6} A_{\alpha}
\simeq 10^{-4}\ (4 \times 10^{-5})
\end{equation}
for $\nu_e - \nu_s$ ($\nu_{\mu,\, \tau} - \nu_s$) oscillations.

Additional bounds arise from evaluating Eq.(\ref{whocares})
alternatively at $\lambda \sim D$, i.e., by requiring that
\begin{equation}
D^2 \gg \left| \frac{D}{\beta} \frac{d \beta}{dt} -
        \frac{d \lambda}{dt} \right|.
\end{equation}
The relative importance of the two terms above is not immediately
obvious by inspection.  Suppose the first term dominates in the
region of interest.  The resulting inequality is but a less severe
version of Eq.(\ref{Tbound}).  Supposing now that the second term
is predominant and that there are no accidental cancellations such
that the inequality roughly reduces to two separate conditions
\begin{eqnarray}
D^2 & \gg & \left| \frac{11 T^2}{m_{Planck}}
    \left[\frac{D}{2} + \frac{\Delta m^2}{2 p}
        \left( 3 b \pm 2 a \right) \right] \right|, \nonumber \\
D^2 & \gg & \left| \frac{11 T^3}{2 m_{Planck}}
    \frac{\Delta m^2}{2p} \frac{a}{L^{(\alpha)}}
        \frac{d L^{(\alpha)}}{dT} \right|.
\end{eqnarray}
While the former is essentially another encrypted bound on the
temperatures at which our approximations are valid, similar to
that obtained earlier in Eq.(\ref{Tbound}), the latter amounts to
demanding that
\begin{equation}
\label{dLdt} \left| \frac{d L^{(\alpha)}}{dT} \right| \ll 3 \times
10^{-12} y_{\alpha}^2 \left( \frac{T}{ \text{MeV}} \right)^{4}
\frac{1}{\text{MeV}}
\end{equation}
at the mean momentum.  This condition agrees remarkably well with
that obtained in Ref.\ \cite{longpaper} and should be checked when
integrating the static approximation equation (\ref{Lev4}) for
self-consistency.

We shall not reproduce here the constraints resulting from the
$({\cal U}\partial {\cal U}^{-1}/\partial t)_{11}$ and $({\cal
U}\partial {\cal U}^{-1}/\partial t)_{22}$ corrections to the
eigenvalues $k_1$ and $k_2$ respectively, as they appear to be
much less stringent than those arising from earlier
considerations. Indeed, this is to be expected given that the
correction terms are only of the order $\beta^2$ where the small
$\beta$ limit is effective.

Turning our attention now to the effects of ${\cal U} \partial {\cal U}^{-1}/\partial t$
on the instantaneous eigenvectors, we observe that, using a
perturbative analysis, the ``first
order'' corrections to $Q_1^{(0)} (t), \, Q_2^{(0)} (t)$ and
$Q_3^{(0)}(t)$ take on the forms
\begin{eqnarray}
Q_1^{(1)}(t) & = &
    \frac{\left( {\cal U} \frac{\partial {\cal U}}{\partial t}^{-1} \right)_{21}}
                {k_1 - k_2} Q_2^{(0)}(t)
    + \frac{ \left( {\cal U} \frac{\partial {\cal U}}{\partial t}^{-1} \right)_{31}}
            {k_1 - k_3} Q_3^{(0)}(t)
\simeq  \frac{ \left( {\cal U} \frac{\partial {\cal U}}{\partial t}^{-1} \right)_{31}}
            {k_1} Q_3^{(0)}(t), \nonumber \\
Q_2^{(1)}(t) & = &
    \frac{\left( {\cal U} \frac{\partial {\cal U}}{\partial t}^{-1} \right)_{12}}
                {k_2 - k_1} Q_1^{(0)}(t)
    + \frac{ \left( {\cal U} \frac{\partial {\cal U}}{\partial t}^{-1} \right)_{32}}
            {k_2 - k_3} Q_3^{(0)}(t)
 \simeq  \frac{\left( {\cal U} \frac{\partial{\cal U}}{\partial t}^{-1} \right)_{32}}
                {k_2} Q_3^{(0)}(t)
\simeq \frac{ \left( {\cal U} \frac{\partial {\cal U}}{\partial t}^{-1} \right)_{31}}
            {k_1} Q_3^{(0)}(t), \nonumber \\
Q_3^{(1)}(t) & = &
    \frac{\left( {\cal U} \frac{\partial {\cal U}}{\partial t}^{-1} \right)_{13}}
                {k_3 - k_1} Q_1^{(0)}(t)
    + \frac{ \left( {\cal U} \frac{\partial {\cal U}}{\partial t}^{-1} \right)_{23}}
            {k_3 - k_2} Q_2^{(0)}(t)    \nonumber \\
& \simeq &
    -\left|\frac{\sqrt{2} \left( {\cal U} \frac{\partial {\cal U}}{\partial t}^{-1} \right)_{13}}
            {k_1} \right|
    \frac{1}{\sqrt{2}} \left[ \exp(i \phi) Q_1^{(0)}(t)
            + \exp(- i \phi) Q_2^{(0)}(t) \right],
\end{eqnarray}
where
\begin{equation}
\exp(i \phi) = \frac{\frac{\left( {\cal U} \frac{\partial {\cal U}}{\partial t}^{-1} \right)_{13}}
        {k_1}}
        {\left| \frac{\left( {\cal U} \frac{\partial {\cal U}}{\partial t}^{-1} \right)_{13}}
        {k_1} \right|}
\end{equation}
and we have neglected terms proportional to $\beta^2$.  Thus the
single requirement that must be satisfied in this case in the
small $\beta$ limit is
\begin{equation}
\left| \frac{ \sqrt{2} \left( {\cal U} \frac{\partial {\cal U}}{\partial t}^{-1} \right)_{13}}
        {k_1} \right| \ll 1.
\end{equation}
It turns out that the above condition does not in fact lead to
more severe bounds than those following from other requirements.

\subsection{Oscillation dominated asymmetry evolution}
\label{lowtemp}

At low temperatures, such that $D \ll |\beta|, |\lambda|$, the collision rate becomes much
slower
and the generation of lepton number is dominated by oscillations.  Taking the $D=0$ limit,
we can demonstrate that the QKEs reduce to usual expression for MSW transitions.

We begin by setting $D=0$ in Eq.(\ref{approxeq}).  In this case, setting
$\partial {\cal U}^{-1}/\partial t = 0$ exactly corresponds to the usual adiabatic
approximation for MSW evolution.
The eigenvalues now become
\begin{equation}
k_1=k_2^*=i\sqrt{\beta^2+\lambda^2}, \hspace{10mm} k_3=0,
\end{equation}
and the matrix ${\cal U}^{-1}(={\cal U}^{\dagger})$ is given by
\begin{equation}
{\cal U}^{-1}=\frac{1}{\sqrt{2(\beta^2+\lambda^2)}}
\left( \begin{array}{ccc}
\lambda\ & \lambda\ & \sqrt{2}\beta \\
-i\sqrt{\beta^2+\lambda^2}\ & i\sqrt{\beta^2+\lambda^2}\ & 0 \\
-\beta\ & -\beta\ & \sqrt{2}\lambda
\end{array} \right).
\end{equation}
With the initial conditions $P_x(0)\simeq0$, $P_y(0) \simeq 0$, the solution
for $P_z(t)$ is given by
\begin{eqnarray}
\label{Pz}
P_z(t) & \simeq & \sum_{i,\gamma} {\cal U}^{-1}_{z i}(t)\exp\left[\int_0^t k_i(t') dt'\right]
{\cal U}_{i \gamma}(0)P_{\gamma }(0) \nonumber \\
& \simeq & \frac{\lambda(t)}{\sqrt{\beta^2(t)+\lambda^2(t)}}
\frac{\lambda(0)}{\sqrt{\beta^2(0)+\lambda^2(0)}}P_z(0) \nonumber \\
& = & c_{2\theta_m}(t)c_{2\theta_m}(0)P_z(0),
\end{eqnarray}
where $c_{2\theta_m}=\cos 2\theta_m$ and
\begin{equation}
\sin^2(2\theta_m) = \frac{s^2}{s^2+(c-b+a)^2}.
\end{equation}
The number density of sterile neutrinos is then given by
\begin{equation}
N_s(t) \simeq N_{\alpha}(0) \frac{1}{2}
\left[1-c_{2\theta_m}(t)c_{2\theta_m}(0)\right]
+ N_s(0)\left\{1-\frac{1}{2}\left[1-c_{2\theta_m}(t)c_{2\theta_m}(0)\right]\right\}.
\end{equation}
Since the adiabatic transition probability for a neutrino propagating
through a medium where the mixing angle changes from $\theta_m(0)$ to
$\theta_m(t)$ is given by
\begin{equation}
\text{Prob} (\nu_{\alpha} \rightarrow \nu_s) = \frac{1}{2}
\left[1-c_{2\theta_m}(t)c_{2\theta_m}(0)\right],
\end{equation}
we have
\begin{equation}
N_s(t) \simeq N_{\alpha}(0)\text{Prob} (\nu_{\alpha} \rightarrow \nu_s)
+ N_s(0) \text{Prob} (\nu_s \rightarrow \nu_s),
\end{equation}
so, as expected, the adiabatic approximation for the QKEs is equivalent to the adiabatic
limit of the MSW effect if we turn off the decohering collisions.

The low temperature equation, used in previous work \cite{Nnufv,bell} to calculate
the growth of asymmetry by determining how quickly the resonance momentum moves
through the distribution, can be derived from the QKEs in the $D=0$ limit.

Since we have $b \simeq 0$, the resonance condition $c \pm a \simeq 0$
can only be attained for either the neutrinos or antineutrinos, depending
on the sign of the asymmetry.
Assuming for definiteness that $L_{\alpha} > 0$, which implies $a >0$, only the
antineutrino resonance will contribute to the growth of lepton
number, so that
\begin{eqnarray}
\frac{dL_{\alpha}}{dT} &=& \frac{d}{dT} \frac{1}{n_{\gamma}}
\int dp \left[ \frac{1}{2}P_0(1+P_z)N^{\text{eq}}
-\frac{1}{2}\overline{P}_0(1+\overline{P}_z)N^{\text{eq}}
\right], \nonumber \\
& \simeq & - \int \frac{dp N^{\text{eq}} \overline{P}_0}{2 n_{\gamma}}
\frac{d\overline{P}_z}{dT},
\end{eqnarray}
where $\frac{d\overline{P}_0}{dT}\simeq 0$ has been assumed. From Eq.(\ref{Pz}) we find
\begin{equation}
\frac{d\overline{P}_z}{dT}= \overline{P}_z(0)
\left(\frac{c-a(0)}{[s^2+(c-a(0))^2]^{1/2}}\right)
\left( \frac{s^2}{[s^2+(c-a(t))^2]^{3/2}} \right)
\frac{a(t)T}{p_{\text{res}}} \frac{d}{dT} \left(\frac{p_{\text{res}}}{T}\right)
\end{equation}
where we have used $a/c=p/p_{\text{res}}$ and
$\frac{da}{dT}= -\frac{aT}{p_{\text{res}}}
\frac{d}{dT} \left( \frac{p_{\text{res}}}{T} \right)$.
This results in
\begin{eqnarray}
\label{int}
\frac{dL_{\alpha}}{dT} &\simeq& -\frac{T}{2n_{\gamma}p_{\text{res}}} \frac{d}{dT}
\left(\frac{p_{\text{res}}}{T}\right) \nonumber \\
&\times& \int^{p_{\text{res}}+\delta
p_{\text{res}}}_{p_{\text{res}}-\delta p_{\text{res}}}dpN^{\text{eq}}\overline{P}_0
\overline{P}_z(0)
\left(\frac{c-a(0)}{[s^2+(c-a(0))^2]^{1/2}}\right)
\left( \frac{a(t)s^2}{[s^2+(c-a(t))^2]^{3/2}} \right),
\end{eqnarray}
where $\delta p_{\text{res}}$ is the resonance width,
since the last term in Eq.(\ref{int}) ensures the integral is strongly peaked about
the resonance momentum.  Averaging the integrand over the width of the resonance
leads to
\begin{equation}
\frac{dL_{\alpha}}{dT} \simeq
\frac{\overline{N}_{\alpha}(p_{\text{res}}+\delta p_{\text{res}}) -
\overline{N}_s(p_{\text{res}}+\delta p_{\text{res}})}{n_{\gamma}}
T \frac{d}{dT} \left(\frac{p_{\text{res}}}{T}\right)
\frac{s^2}{s^2+(c\delta p_{\text{res}}/p_{\text{res}})^2},
\end{equation}
so that in the limit of zero resonance width we recover
\begin{equation}
\frac{dL_{\alpha}}{dT} \simeq
\frac{\overline{N}_{\alpha} - \overline{N}_s}{n_{\gamma}}
T \frac{d}{dT}\left(\frac{p_{\text{res}}}{T}\right),
\label{dpresdT}
\end{equation}
where the $\overline{N}_{\alpha} - \overline{N}_s$ term is interpreted as evaluated just
before the resonance passes a particular momentum value.  This term expresses the difference
between the number of $\overline{\nu}_{\alpha}$ and $\overline{\nu}_s$ which are
converted at the
resonance, while the factor $\frac{d}{dT} \left( \frac{p_{\text{res}}}{T} \right)$ accounts for how
quickly the resonance momentum moves through the distribution converting active
to sterile neutrinos. See Refs.\cite{Nnufv,bell} for applications of this type of
equation. The derivation given above for Eq.(\ref{dpresdT}) is the first justification
of it from first principles.

\section{Mirror neutrinos instead of sterile neutrinos}
\label{mirror}

We now briefly discuss how the foregoing must be altered to deal with mirror instead of
strictly sterile neutrinos. Mirror neutrinos form the neutral lepton sector of the
mirror fields postulated in the Exact Parity Model (EPM) \cite{epm}. The EPM restores
exact parity
invariance to the microscopic world by parity-doubling the Standard Model.
Interestingly, if neutrinos and mirror neutrinos have mass, and if the two sector mix,
then in the absence of intergenerational mixing the mass eigenstates are maximal
superpositions of ordinary and mirror neutrinos.  Maximal oscillations of $\nu_{\mu}$
with its mirror partner $\nu'_{\mu}$ can explain the atmospheric neutrino anomaly,
while maximal oscillation of $\nu_e$ with {\it its} mirror partner $\nu'_e$ can
similarly solve the solar neutrino problem. The maximal mixing of $\nu_{\alpha}$ with
$\nu'_{\alpha}$ is a consequence of the unbroken parity symmetry.  This form of parity
symmetry is arguably one of the most credible explanations for the maximal mixing of
muon-neutrinos observed in atmospheric neutrino experiments. The LSND anomaly can be
accomodated by switching on small intergenerational mixing with the appropriate mass
hierarchy.

The issue of whether or not the EPM explanation of the neutrino anomalies is consistent
with Big Bang Nucleosynthesis is an important one. We assume that in the very early
stages of the Big Bang ordinary matter predominates over mirror matter. (See
Refs.\cite{kolb} for
speculations about how this could arise.) The question is then whether interactions
between the ordinary and mirror sectors at later times overcreates mirror matter,
spoiling BBN. Focussing on ordinary-mirror neutrino mixing, we note that a sufficiently
large ($\stackrel{>}{\sim} 10^{-5}$) pre-existing neutrino asymmetry can always be
invoked to suppress ordinary-mirror oscillations, thus leading to standard BBN despite
the presence of the mirror sector \cite{prl}. However, it is more interesting to suppose
that
the neutrino sector of the EPM ``saves itself'' through the production of large neutrino
asymmetries in the manner discussed above. In order to do this analysis, the QKEs have
to be modified to take into account the interactions of mirror neutrinos amongst
themselves and with other mirror species such as the mirror photon and the mirror
electron. See Ref.\cite{astropart} for the proof that the EPM is consistent with BBN in
an interesting region of parameter space.

Mirror neutrinos feel mirror weak interactions. The interaction strength is given by
$G_F$, since exact parity symmetry imposes equality of coupling constants and gauge
boson masses between the sectors. We may classify the additional terms in the QKEs
through powers of $G_F$: coherent forward scattering induces $O(G_F)$ terms, while
incoherent scattering induces terms of $O(G_F^2)$ and higher. Provided that the number
densities of mirror neutrinos remains low, {\it all terms other than the $O(G_F)$
coherent forward scattering effects can be neglected.} If the number densities of mirror
neutrinos become appreciable, then many of the complexities inherent in active-active
scattering become an issue.

So, provided one is working in a parameter space region where the mirror neutrino
number densities remain small, it is consistent to alter the QKEs of the strictly
sterile neutrino case by the simple substitution,
\begin{equation}
L^{(\alpha)} \to L^{(\alpha)} - L'^{(\beta)},
\end{equation}
for $\nu_{\alpha} - \nu'_{\beta}$ oscillations ($\alpha,\beta = e,\mu,\tau$). The
quantity $L'^{(\beta)}$ is simply the mirror version of $L^{(\beta)}$. This substitution
fully incorporates the coherent forward scattering of mirror neutrinos off mirror neutrinos
induced by mirror weak interactions.

It is important to understand why these $O(G_F)$ mirror neutrino self-interactions are
as important as $O(G_F)$ neutrino self-interactions. The term $a(p)$ in the effective
matter potential is equal to the sum of terms directly proportional to the difference in
number densities between neutrinos and antineutrinos, and the difference in number
densities between mirror neutrinos and mirror antineutrinos.  These number density
differences will be of
the same order of magnitude simply because mirror neutrinos and antineutrinos are
produced, via partially incoherent oscillations, from ordinary neutrinos and
antineutrinos. Since we know that the $a(p)$ term dominates the $b(p)$ term once
significant neutrino asymmetries have been created, and since coherent effects anyway
become more and more important as the temperature approaches the BBN epoch, the $O(G_F)$
mirror neutrino self-interactions are of critical importance. The fact that they are
completely non-negligible distinguishes the cosmology of mirror neutrinos from that of
strictly sterile neutrinos \cite{forthcoming}.

\section{Multi-flavour oscillations}
\label{main2}

The purpose of this section is to begin the study of multi-flavour effects.  In
particular, we wish to determine the conditions under which it is appropriate to break
up a multi-flavour situation into effective two-flavour subsystems, where two-flavour
expressions can be applied.  An understanding of when multi-flavour effects may arise is
important, because while two-flavour subsystems are a useful approximation, a realistic
situation will always involve three or more flavours.  A multi-flavour system may
involve features that do not occur in simpler two-flavour systems. We would expect, for
instance, genuine multi-flavour effects to arise in systems in which there are
``overlapping'' resonances.  With a general multi-flavour system, there is also the
possibility of explicit $CP$ violation due to complex phases in the neutrino mixing
matrix.

Partially incoherent oscillations in a multi-flavour system have not previously been
studied from first principles. In all prior work on lepton asymmetries, multi-flavour
systems were dealt with in terms of two-flavour subsystems.

The system we shall consider here, chosen both for definiteness and importance, consists
of $\nu_{\mu}$ maximally mixed a sterile neutrino $\nu_s$, and a heavier $\nu_{\tau}$
which has small mixing with $\nu_{\mu}$ and $\nu_s$.  This particular system is well
motivated by the atmospheric neutrino anomaly, with the parameter space region $\sin
2\theta_{\mu s}\simeq 1$ and $\Delta m^2_{\mu s}\sim 10^{-2}-10^{-3} \text{eV}^2$
leading to a resolution of the anomaly via $\nu_{\mu} \to \nu_s$ oscillations. The
$\nu_{\tau,\mu,s}$ system was studied in Ref.\cite{longpaper}, where the role of the
heavier tau neutrino, taken to have a mass in the range where $\Delta m^2_{\tau s}=
1-1000\text{eV}^2$, is to generate an $L_{\tau}$ asymmetry via oscillations with the
sterile neutrino as per the dynamics discussed in the previous Section.  The $L_{\tau}$
asymmetry generates a large Wolfenstein term in the effective potential for the
$\nu_{\mu} - \nu_s$ subsystem, which then acts to suppress oscillations between the muon
and sterile neutrinos. For a range of parameters, this prevents the maximal
$\nu_{\mu}-\nu_s$ oscillations from bringing the sterile neutrinos into equilibrium, thus
circumventing the supposedly stringent BBN bounds on the mixing of sterile and active
neutrinos. {\it This phenomenon is vital for reconciling the $\nu_{\mu} \to \nu_s$
solution to the atmospheric neutrino problem with cosmology.} Note, however, that in
previous analyses the $\nu_{\tau} -\nu_s$ and $\nu_{\mu} - \nu_s$ resonances were dealt
with in a two-flavour manner;  possible three-flavour effects were ignored. Given that
the resonances are well separated, this is expected to be a good approximation. However,
in order to verify this is the case, we need to study it within a full three-flavour
framework.

We begin by writing down the evolution equations for a three-flavour partially incoherent
system using the density matrix formalism.  While it is in a sense straightforward to
generalise the Quantum Kinetic Equations to a three-flavour system, qualitatively
different complexities are encountered due to the presence of active-active neutrino
oscillations in addition to the more readily understood active-sterile oscillations.
These active-active complications result in equations from which it is difficult to
obtain analytical insight.\footnote{See Ref.\cite{mckellar} for details of the (momentum
dependent) Quantum Kinetic Equations for active-active oscillations.} However, under the
momentum averaged approximation the equations simplify significantly, allowing progress
to be made.  Although going to the momentum averaged approximation will introduce some
errors, it is probably a reasonable approach. In any case, we adopt it here because
it makes possible a first attempt at a first-principles analysis of a three-flavour
system. (The differences between including and excluding the momentum spread for
two-flavour active-sterile oscillations were discussed at length in
Ref.\cite{longpaper}. The qualitative features of the evolution were found to be
preserved in the mean momentum approximation.)

The three-flavour momentum averaged density matrix $\langle\rho\rangle$ is parameterised
in terms of
the SU(3) Gell-Mann matrices $\lambda_i$, $i=1, \ldots, 8$ such that
\begin{equation}
\langle\rho\rangle = \frac{1}{2} P_0 (1+\lambda _i P_i),
\end{equation}
or
\begin{eqnarray}
\label{rho}
\left( \begin{array}{ccc}
\langle\rho\rangle_{\tau \tau} & \langle\rho\rangle_{\tau s} &
\langle\rho\rangle_{\tau \mu} \\
\langle\rho\rangle_{s \tau } & \langle\rho\rangle_{ss} & \langle\rho\rangle_{s \mu}\\
\langle\rho\rangle_{\mu \tau} & \langle\rho\rangle_{\mu s} & \langle\rho\rangle_{\mu \mu}
\end{array} \right)
=\frac{1}{2}P_0 \left( \begin{array}{ccc}
1+P_3+\frac{1}{\sqrt{3}}P_8 & P_1-iP_2 & P_4-iP_5 \\
P_1+iP_2 & 1-P_3+\frac{1}{\sqrt{3}}P_8  & P_6-iP_7 \\
P_4+iP_5 & P_6+iP_7 & 1-\frac{2}{\sqrt{3}}P_8
\end{array}\right).
\end{eqnarray}
The neutrino number densities are related to the diagonal entries
of $\langle\rho\rangle$:
\begin{eqnarray}
n_{\nu_{\tau}}&=&\frac{1}{2}P_0(1+P_3+\frac{1}{\sqrt{3}}P_8)n^{\text{eq}}, \nonumber \\
n_{\nu_s}&=&\frac{1}{2}P_0(1-P_3+\frac{1}{\sqrt{3}}P_8)n^{\text{eq}}, \nonumber \\
n_{\nu_{\mu}}&=&\frac{1}{2}P_0(1-\frac{2}{\sqrt{3}}P_8)n^{\text{eq}}.
\end{eqnarray}
The evolution of a momentum averaged density matrix is described by the
Quantum Rate Equations (QREs) rather than the Quantum Kinetic Equations.
Generalising the two-flavour QREs derived in Ref.\cite{mckellar},  which involves
the lengthy but straightforward procedure of substituting Eq.(\ref{rho}) into the
general QKE expression given by Eq.(22) of Ref.\cite{mckellar},
we find the three-flavour QREs
describing the considered system to be,

\begin{eqnarray}
\label{QRE}
\frac{d}{dt} {\bf P} &=& {\bf V} \times {\bf P}
- D( P_1 \hat{x}_1 + P_2 \hat{x}_2 + P_6 \hat{x}_6 + P_7 \hat{x}_7 )
- D'( P_4 \hat{x}_4 +P_5 \hat{x}_5 ) \nonumber \\
&-& C( \overline{P}_4 \hat{x}_4 -\overline{P}_5
\hat{x}_5 )
-( P_1 \hat{x}_1 + P_2 \hat{x}_2 + P_4 \hat{x}_4
+P_5\hat{x}_5 + P_6 \hat{x}_6 + P_7 \hat{x}_7 )\frac{d}{dt}\ln P_0 \nonumber \\
&+&\frac{2}{3}\left[\left(\frac{3}{2}-P_3\right)\frac{R_{\tau}}{P_0}-P_3\frac{R_{\mu}}{P_0}
\right]\hat{x}_3
+\frac{2}{3}\left[\left(\frac{\sqrt{3}}{2}-
P_8\right)\frac{R_{\tau}}{P_0}-\left(\sqrt{3}+P_8\right)
\frac{R_{\mu}}{P_0}\right]\hat{x}_8\nonumber \\
&+&(-P_6\text{Re}H-P_7\text{Im}H)\hat{x}_1+(-P_6\text{Im}H+P_7\text{Re}H)\hat{x}_2 \nonumber \\
&+&(-P_1\text{Re}H-P_2\text{Im}H)\hat{x}_6+(-P_1\text{Im}H+P_2\text{Re}H)\hat{x}_7,
\end{eqnarray}
with the evolution of $P_0$, which is related to the total number density
of $\nu_{\tau}$, $\nu_{\mu}$ and $\nu_s$, given by
\begin{equation}
\frac{dP_0}{dt}=\frac{2}{3}(R_{\tau}+R_{\mu}),
\end{equation}
where $R_{\alpha}$ is the repopulation function \cite{mckellar}
\begin{equation}
R_{\alpha}= \frac{1}{(n^{\text{eq}})^2} \sum_{j=e,\nu_e,\nu_{\mu},\nu_{\tau}}
\langle\Gamma(j \overline{j} \rightarrow \nu_{\alpha}\overline{\nu}_{\alpha})\rangle
 \left[ h_j n_jn_{\overline{j}}- n_{\nu_{\alpha}} n_{\overline{\nu}_{\alpha}} \right]
-\frac{1}{2}\sum_{i=e,\nu_e} G_i \left[ P_4\overline{P}_4-P_5\overline{P}_5 \right],
\end{equation}
where $h_e=\frac{1}{4}$, $h_{\nu}=1$ and the values for the collision rates
$\langle\Gamma(j \overline{j} \rightarrow \nu_{\alpha}\overline{\nu}_{\alpha})\rangle$
can be found in Refs.\cite{mckellar,et}.  The $n$'s are normalised
such that $n_{\nu_{\alpha}}^{\text{eq}}=n^{\text{eq}}$ and
$n_e^{\text{eq}}=2n^{\text{eq}}$.
The $G$ terms arise from the mixed active-active part of the density matrix,
with $G_e=0.26G_F^2T^5$ and $G_{\nu_e}=0.51G_F^2T^5$.

The quantity
$D=\frac{1}{2}\langle \Gamma_{\alpha} \rangle$ is the damping parameter,
with $\Gamma_{\alpha}$
as given by Eq.(\ref{Gammma_ave}), while $D'$ is the
equivalent parameter for the active-active $\nu_{\tau}-\nu_{\mu}$ oscillations:
\begin{equation}
D' \simeq 1.2G_F^2T^5.
\end{equation}
The parameter $C$ in Eq.(\ref{QRE}) is a damping-like term which couples the 
neutrino and antineutrino density matrices through the off-diagonal $\rho_{\mu \tau}$ 
and $\rho_{\overline{\mu} \overline{\tau}}$ elements, and only arises in the case of mixed 
active-active neutrinos.
For $\nu_{\tau}-\nu_{\mu}$ oscillations \cite{mckellar},
\begin{equation}
C \simeq 1.8G_F^2T^5.
\end{equation}
The quantity $H$ is given by
\begin{eqnarray}
\label{H}
H & = &  \frac{\langle \rho_{\overline{\tau} \overline{\mu}}\rangle \pi}{n^{\text{eq}}}
\int dk'dp'dpdk
\delta_E(k+p-k'-p') f(k)f(p)  \nonumber \\
& \times & \sum_j [V(\nu_{\mu}(k),\overline{\nu}_{\mu}(p)|j(k'),\overline{j}(p'))]
[V(j(k'),\overline{j}(p')|\nu_{\tau}(k),\overline{\nu}_{\tau}(p))],
\end{eqnarray}
where $\int dp \equiv \frac{1}{(2\pi)^3} \int d^3p$.  This is an order $G_F^2T^5$ quantity,
however it will be small compared to $D$ and $C$, provided
that $\rho_{\overline{\tau} \overline{\mu}}$ is small.

We define
\begin{equation}
({\bf V} \times {\bf P})^k \equiv V^iP^jf^{ijk},
\end{equation}
where $f^{ijk}$ are the $SU(3)$ structure constants.
Finally, ${\bf V}$ is given by
\begin{eqnarray}
{\bf V}&=& 2\text{Re} E^{\tau s} \hat{x}_1 - 2\text{Im} E^{\tau s} \hat{x}_2 +
(E^{\tau\tau}-E^{ss}) \hat{x}_3
+ 2\text{Re} E^{\tau \mu} \hat{x}_4 - 2\text{Im} E^{\tau \mu} \hat{x}_5 \nonumber \\
&+& 2\text{Re} E^{s \mu} \hat{x}_6 - 2\text{Im} E^{s \mu} \hat{x}_7
+ \frac{1}{\sqrt{3}}(E^{\tau\tau}+E^{ss}-2E^{\mu \mu}) \hat{x}_8,
\end{eqnarray}
with
\begin{equation}
E^{\alpha \beta}= \omega^{\alpha \beta}+ V^{\alpha \beta},
\end{equation}
where  $\omega^{\alpha \beta}$ are the (vacuum) energy eigenvalues in
the flavour basis,
\begin{equation}
\omega^{\alpha \beta}=\frac{1}{2p}U \text{diag}(m_1^2,m_2^2,m_3^2) U^{\dagger},
\end{equation}
with $U$ being the mixing matrix relating the neutrino mass and flavour
eigenstates. $V^{\alpha \beta}$ are the effective potential terms which,
as well as the usual diagonal potential terms
($V^{\alpha\alpha}=\frac{\Delta m^2}{2p}(-a+b)$), also include
off-diagonal contributions given by
\begin{equation}
V^{\mu\tau}= \frac{\sqrt{2}G_Fn_{\gamma}}{8}
\left[ (\rho_{\mu\tau}-\rho_{\overline{\mu}\overline{\tau}})
- \frac{14\zeta(4)}{\zeta(3)}\frac{2T}{3M^2_Z}\rho_{\mu\tau} \right].
\end{equation}

The appearance of off-diagonal terms in the effective potential is a consequence
of active-active oscillations.  They depend on the quantities $P_4$ and $P_5$ which
parameterise the mixed $\nu_{\tau}$ and $\nu_{\mu}$ states in the density matrix.
Note, however, that the number densities of $\nu_{\tau}$ and $\nu_{\mu}$ are
equal (to order $L$).
As we shall see, this results in $P_4$ and $P_5$ being suppressed, with
respect to $P_1$ \& $P_2$ and $P_6$ \& $P_7$ which parameterise the mixed
$\nu_{\tau,s}$ and $\nu_{\mu,s}$ subsystems, respectively.

For simplicity, the $\nu_{\tau}-\nu_{\mu}$ mixing angle will be set to zero.
Observe, however, that there will still be a small effective
mixing between $\nu_{\tau}$ and $\nu_{\mu}$ indirectly through $\nu_s$.  We parameterise
the mixing matrix $U$ as
\begin{eqnarray}
U=\left( \begin{array}{ccc}
c_{\phi} & s_{\phi} & 0 \\
-s_{\phi} & c_{\phi} & 0 \\
0 & 0 & 1
\end{array}\right)
\left( \begin{array}{ccc}
1 & 0 & 0 \\
0 & \frac{1}{\sqrt{2}} & \frac{1}{\sqrt{2}} \\
0 & -\frac{1}{\sqrt{2}} & \frac{1}{\sqrt{2}}
\end{array}\right),
\end{eqnarray}
where $\phi$ is the $\nu_{\tau}-\nu_s$ mixing angle, and the
$\nu_{\mu}-\nu_s$
mixing angle has been fixed at $\frac{\pi}{4}$.

Unlike the two-flavour case, the neutrino and anti-neutrino density matrices are
coupled through the $C$ term in Eq.(\ref{QRE}), so that we must consider a system
of 16 coupled differential equations for $P_1,
\ldots, P_8,\overline{P}_1,\ldots,\overline{P}_8$.
As in the two-flavour case, we adopt the $\frac{dP_0}{dt} \simeq 0$ approximation.
This allows Eq.(\ref{QRE}) to be expressed in the form:
\begin{equation}
\frac{d}{d t} {\bf P} = {\cal K} {\bf P},
\end{equation}
where ${\bf P}=(P_1,...P_8,\overline{P}_1,...,\overline{P}_8)$ and
${\cal K}$ is the matrix given by
\begin{eqnarray}
\label{blockmatrix}
\cal{K}=
\left( \begin{array}{cc}
\cal{M} & \cal{C} \\
\cal{C} & \overline{\cal{M}}
\end{array} \right),
\end{eqnarray}
where $\cal{M}$ and $\cal{C}$ are the submatrices
\begin{eqnarray}
\label{matrix}
\cal{M}=
\left(\begin{array}{cccccccc}
-D                &-\lambda         &0                 & 0
&\frac{1}{2}\gamma&\frac{1}{2}\delta_I -H_R  &\frac{1}{2}\delta_R -H_I & 0 \\
\lambda           &-D               &-\beta &-\frac{1}{2}\gamma
&0                &\frac{1}{2}\delta_R -H_I & -\frac{1}{2}\delta_I +H_R  & 0 \\
0                 &\beta            &0      &\frac{1}{2}\delta_I
&\frac{1}{2}\delta_R &0      &-\frac{1}{2}\gamma & 0 \\
0                 &\frac{1}{2}\gamma&-\frac{1}{2}\delta_I      &-D'
&-\sigma     &0       &-\frac{1}{2}\beta &-\frac{\sqrt{3}}{2}\delta_I \\
-\frac{1}{2}\gamma& 0               & -\frac{1}{2}\delta_R & \sigma
& -D'             &\frac{1}{2}\beta & 0 & -\frac{\sqrt{3}}{2}\delta_R \\
-\frac{1}{2}\delta_I-H_R  & -\frac{1}{2}\delta_R-H_I & 0 & 0
&-\frac{1}{2}\beta& -D              & -\epsilon  & 0 \\
-\frac{1}{2}\delta_R-H_I &\frac{1}{2}\delta_I+H_R  &\frac{1}{2}\gamma  & \frac{1}{2}\beta
& 0               & \epsilon        & -D & -\frac{\sqrt{3}}{2}\gamma  \\
0    & 0   & 0    & \frac{\sqrt{3}}{2}\delta_I
& \frac{\sqrt{3}}{2}\delta_R  & 0 & \frac{\sqrt{3}}{2}\gamma     & 0
\end{array} \right),
\end{eqnarray}

\begin{eqnarray}
\label{cmatrix}
\cal{C}=
\left(\begin{array}{cccccccc}
0 & 0 & 0 & 0 & 0 & 0 & 0 & 0 \\
0 & 0 & 0 & 0 & 0 & 0 & 0 & 0 \\
0 & 0 & 0 & 0 & 0 & 0 & 0 & 0 \\
0 & 0 & 0 & C & 0 & 0 & 0 & 0 \\
0 & 0 & 0 & 0 & -C & 0 & 0 & 0 \\
0 & 0 & 0 & 0 & 0 & 0 & 0 & 0 \\
0 & 0 & 0 & 0 & 0 & 0 & 0 & 0 \\
0 & 0 & 0 & 0 & 0 & 0 & 0 & 0
\end{array} \right),
\end{eqnarray}
and $\overline{\cal{M}}$ is obtained from $\cal{M}$ via the replacement
$\lambda \rightarrow \overline{\lambda}$, $\sigma \rightarrow \overline{\sigma}$,
$\epsilon \rightarrow \overline{\epsilon}$, $D \rightarrow \overline{D}$,
$D' \rightarrow \overline{D}'$, and $H \rightarrow \overline{H}$.
We use the notation, $\delta_I=\text{Im}\delta$, $\delta_R=\text{Re}\delta$,
$H_I=\text{Im}H$, $H_R=\text{Re}H$, and
\begin{eqnarray}
\lambda &=& V^{\tau} - \frac{\Delta m^2}{2p} c_{2\phi}(1-\frac{r}{2}), \nonumber \\
\sigma &=&V^{\tau}-V^{\mu}- \frac{\Delta m^2}{2p}c^2_{\phi}(1-\frac{r}{2}), \nonumber \\
\epsilon&=&-V^{\mu} - \frac{\Delta m^2}{2p} s^2_{\phi}(1-\frac{r}{2}),
\nonumber \\
\beta &=&  \frac{\Delta m^2}{2p} s_{2\phi}(1-\frac{r}{2}), \nonumber \\
\delta&=& V^{\tau\mu} - \frac{\Delta m_{\mu s}^2}{2p} s_{\phi}, \nonumber \\
\gamma&=& -\frac{\Delta m_{\mu s}^2}{2p} c_{\phi}, \nonumber \\
\end{eqnarray}
where $\Delta m^2 = \Delta m^2_{\tau s}$ and
$r = \Delta m^2_{\mu s} / \Delta m^2_{\tau s} \ll 1$.
The parameters
$\beta$ and $\lambda$ take roughly the same form as for the two-flavour active-sterile
system, and are related to the $\nu_{\tau}-\nu_s$ oscillations. The quantities
$\delta$ and $\gamma$ are the equivalent of $\beta$ for the
$\nu_{\tau}-\nu_{\mu}$ and $\nu_{\mu}-\nu_s$ oscillations, respectively,
while $\sigma$ and $\epsilon$ are the $\nu_{\tau}-\nu_{\mu}$ and $\nu_{\mu}-\nu_s$
analogues of $\lambda$.

From Eq.(\ref{QRE}) and the approximations adopted as per the discussion above, the
evolution of the lepton numbers $L_{\tau}$ and $L_{\mu}$ are found to be
\begin{eqnarray}
\frac{dL_{\tau}}{dt}=
\frac{ n^{\text{eq}} }{2n_{\gamma}}\left[P_0(\beta P_2 + \delta_I P_4+ \delta_R P_5 ) -
\overline{P}_0(\beta \overline{P}_2++ \delta_I \overline{P}_4+ \delta_R
\overline{P}_5)\right],  \nonumber \\
\frac{dL_{\mu}}{dt}=
\frac{n^{\text{eq}}}{2n_{\gamma}}\left[-P_0(\gamma P_7 + + \delta_I P_4+\delta_R P_5) +
\overline{P}_0(\gamma \overline{P}_7++ \delta_I \overline{P}_4+ \delta_R
\overline{P}_5)\right].
\end{eqnarray}
To solve these equations, $P_2$, $P_4$, $P_5$ and $P_7$ must be determined
as functions of time.

A first inspection of Eqs.(\ref{blockmatrix}), (\ref{matrix}) and (\ref{cmatrix}) gives
the impression that the
various oscillation modes are coupled together in a non-trivial
manner.  However, under the assumptions made, the solution reveals that
the three oscillation modes effectively decouple to first
order in the small parameters $\beta$, $\delta$, $\gamma$ and $H$.

To solve these sixteen coupled equations, the matrix will be
diagonalised in the small $\beta,\delta,\gamma,H$ limit, in analogy to the
two-flavour case.  This is done by treating the $\beta,\delta,\gamma,H$ terms as
a perturbation to the zeroth order
solution obtained with $\beta=\delta=\gamma=H=0$.
The parameters $\beta$ and $\gamma$ may be of approximately the same order of magnitude,
depending on the mass-squared differences and mixing angles.
The size of $\delta$ depends on the size of the off-diagonal potential term $V^{\tau\mu}$
which, as with $H$, is dependent on the values of $P_4$ and $P_5$.
Our perturbative solution relies on the assumption that $P_4$ and $P_5$ are small. This
is expected for the physical reason of decoherent damping through $D'$, except at low
temperatures (say in the neutrino decoupling regime).
The solution obtained with this assumption may then be checked for self-consistency.

For the purposes of the perturbation, we treat
$\beta,\delta,\gamma,H$ as being of the same order.  Again, as with the
two-flavour case, we adopt the adiabatic-like approximation that ${\cal U}
\frac{\partial
{\cal U}}{\partial t}^{-1} \simeq 0$ where ${\cal U}$ is the time dependent
matrix which instantaneously diagonalises the matrix in Eq.(\ref{blockmatrix}).
Solving for the appropriate eigenvectors to first order in $\beta,\delta,\gamma,H$
yields
\begin{eqnarray}
\label{p3flav}
P_2(t)&=& -\frac{\beta D}{D^2+\lambda^2} \left( \frac{n_{\nu_{\tau}}-n_{\nu_s}}{P_0
n^{\text{eq}}} \right), \nonumber \\
P_4(t)&=& Y_4 \left(\frac{n_{\nu_{\tau}}-n_{\nu_{\mu}}}{P_0 n^{\text{eq}}} \right)
+ Z_4 \left(\frac{n_{\overline{\nu}_\tau}-
n_{\overline{\nu}_\mu}}{\overline{P}_0 n^{\text{eq}}} \right), \nonumber \\
P_5(t)&=& Y_5 \left(\frac{n_{\nu_{\tau}}-n_{\nu_{\mu}}}{P_0 n^{\text{eq}}} \right)
+ Z_5 \left(\frac{n_{\overline{\nu}_\tau}-
n_{\overline{\nu}_\mu}}{\overline{P}_0 n^{\text{eq}}} \right), \nonumber \\
P_7(t)&=& \frac{\gamma D}{D^2+\epsilon^2} \left( \frac{n_{\nu_{\mu}}-n_{\nu_s}}{P_0
n^{\text{eq}}} \right),
\end{eqnarray}
where $Y_{4,5}$, $Z_{4,5}$ are order $\delta$ terms and are functions of the parameters
$D'$, $\sigma$, $\overline{\sigma}$ and $C$, arising from the diagonalisation of
the $P_4,P_5,\overline{P}_4,\overline{P}_5$ submatrix
\begin{eqnarray}
\left(\begin{array}{cccc}
-D' & -\sigma & C & 0 \\
\sigma & -D' & 0 & -C \\
C & 0 & -\overline{D}' & -\overline{\sigma} \\
0 & -C & \overline{\sigma} & -\overline{D}'
\end{array}\right).
\end{eqnarray}

We see that $P_5$ (and similarly $P_4$) is suppressed not only because $\delta$ is small but
also since $n_{\nu_{\tau}}-n_{\nu_{\mu}} = O(L)$.  This ensures that $V^{\tau \mu}$
and $H$ remain small, consistent with the assumptions made.
The proportionality of $P_4$ and $P_5$ to $n_{\nu_{\mu}} -
n_{\nu_{\tau}}$ is one manifestation of the unimportance of oscillations between species of
roughly equal number densities.

These expressions in Eq.(\ref{p3flav}) resemble those that would be obtained by 
breaking the system down 
into two two-flavour subsystems $\nu_{\tau} + \nu_s$ and $\nu_{\mu} + \nu_s$.
The evolution equations for $L_{\tau}$ and
$L_{\mu}$ are coupled only via the effective potentials and $n_{\nu_s}$.
In the initial stages
of lepton number creation, when the sterile neutrino number density can be neglected,
the equations are in essence coupled only through the $L$-dependent effective
potentials. The equations for lepton number become
\begin{eqnarray}
\label{L3flav}
\frac{dL_{\tau}}{dt}=
\frac{1}{2n_{\gamma}}\left[ -\frac{\beta^2 D}{D^2+\lambda^2}(n_{\nu_{\tau}}-n_{\nu_s})
+\frac{\beta^2 \overline{D}}{\overline{D}^2
+\overline{\lambda}^2}(n_{\overline{\nu}_{\tau}}-n_{\overline{\nu}_s})
\right],  \nonumber \\
\frac{dL_{\mu}}{dt}=
\frac{1}{2n_{\gamma}}\left[ -\frac{\gamma^2 D}{D^2+\epsilon^2}(n_{\nu_{\mu}}-n_{\nu_s})
+\frac{\gamma^2 \overline{D}}{\overline{D}^2
+\overline{\epsilon}^2}(n_{\overline{\nu}_{\mu}}-n_{\overline{\nu}_s})  \right],
\end{eqnarray}
with the $P_4$ and $P_5$ terms now neglected. These types of equations, generalised in
the obvious way to incorporate the thermal momentum distribution, have been used in
practical calculations \cite{longpaper,astropart}. The above analysis is the first to
fully justify their use on the basis of first principles (admittedly in the Quantum Rate
Equation approximation).

The various oscillation modes would seem only to be coupled via higher
order terms in $\beta,\gamma,\delta,H$.  In fact the restrictions we imposed
on the mixing angles can be relaxed, and the only difference between
Eq.(\ref{L3flav}) and the two-flavour equations used in
Refs.\cite{longpaper,astropart} are the small corrections (in terms of the mass-squared 
differences and mixing angles) to the parameters
$\beta$, $\lambda$, $\gamma$ and $\epsilon$.

This two-flavour subsystem solution will not be valid when the resonances of 
different subsystems are close enough to overlap.  
For example, the $\nu_{\tau} - \nu_s$ and $\nu_{\mu} - \nu_s$ resonances begin 
to overlap if $\lambda \simeq \epsilon$.  In this case some of the eigenvalues 
of the matrix in Eq.(\ref{matrix}) become degenerate and hence the perturbative method 
used to obtain the solution may break down.

The study of the $\nu_{\tau,\mu,s}$ system, with $\nu_{\mu} \to \nu_s$ parameters set by
the atmospheric neutrino anomaly, has recently been re-examined by Foot \cite{newfoot}.
In this work, the ``pairwise two-flavour approximation'' is used, but with each
two-flavour active-sterile subsystem treated using the full two-flavour QKEs.

\section{Conclusion}
\label{conc}

Active-sterile neutrino oscillations will play an important role in early universe
cosmology if light sterile neutrinos exist. The combined solar neutrino, atmospheric
neutrino and LSND anomalies require at least one light sterile flavour. The partially
incoherent neutrino oscillations that occur in the early universe are described using
the Quantum Kinetic Equations. These equations provide an in-principle means of
calculating the cosmological consequences of light sterile neutrinos, with Big Bang
Nucleosynthesis being an important concern. However, their qualitative physical
consequences are not always easy to extract, and they are computationally demanding.
Approximation schemes are therefore welcome.

The main insights of this paper are twofold:
\begin{enumerate}
\item An adiabatic approximation for the partially incoherent oscillations described by
the QKEs has been developed. This clarifies the origin of the ``static approximation''
discussed in Ref.\cite{longpaper}, revealing it to be conceptually related to the
usual adiabatic approximation of matter-affected oscillations. In the absence of
collisions, we have explicitly shown that it reduces to the usual adiabatic
approximation. When collisions are important, we have used it to rederive some very
useful approximate evolution equations for lepton number in the small $\beta$ limit. The
evolution of lepton
number is of particular importance because of the central role played by the Wolfenstein
term in the effective matter potential in suppressing sterile neutrino production prior
to BBN. The systematic approach to approximating the QKEs introduced in this paper
suggests further development which could improve, in a hopefully practical way, on the
adiabatic approximation for partially incoherent oscillations.
\item A first principles treatment of partially incoherent three-flavour oscillations
has been attempted for the first time.
Focussing on a $\nu_{\tau,\mu,s}$ subsystem with maximally mixed $\nu_{\mu}$ and
$\nu_s$, important for resolving the atmospheric neutrino anomaly, the
three-flavour Quantum Rate Equations were explicitly written down. The QREs are obtained
from the QKEs by approximating the evolution of the neutrino momentum distribution by
the evolution of neutrinos having the mean momentum. By employing a similar
adiabatic-like approximation in the small $\beta,\gamma,\delta,H$ limit (three-flavour
static approximation) to this case, it was demonstrated that the
three-flavour system separated into two two-flavour subsystems, $\nu_{\tau} + \nu_s$
and $\nu_{\mu} + \nu_s$, coupled only through the dependence on the Wolfenstein term on
family lepton numbers, and through the common $\nu_s$ ensemble. The ``coupled
two-flavour subsystem'' approach had previously been used in studies which concluded
that the maximal $\nu_{\mu} \to \nu_s$ solution to the atmospheric neutrino anomaly was
consistent with BBN provided that the $\nu_{\tau} - \nu_s$ oscillation parameters lay
in a particular region. The conclusion that there is no cosmological objection to the
$\nu_{\mu} \to \nu_s$ solution to the atmospheric anomaly can thus be made with even
more confidence.
\end{enumerate}

An exciting new era in fundamental physics has begun, ushered in by the beautiful
atmospheric neutrino results of SuperKamiokande. We await with great interest news from
the Sudbury Neutrino Observatory \cite{sno}, and further data from SuperKamiokande,
regarding the
existence or otherwise of light sterile neutrinos. Should they exist, then a new synergy
between the microscopic and macroscopic worlds will be revealed through the evolution of
partially coherent active-sterile (or ordinary-mirror) neutrino oscillations.

\acknowledgments{We would like to thank Robert Foot for his many insights and for
suggesting improvements to an earlier draft of this paper.
RRV is supported by the Australian Research
Council. NFB and YYYW are supported by the Commonwealth of Australia and The University
of Melbourne.}

\newpage

\appendix

\section{Calculation of decoherence funtion}

In this section the calculation of the decoherence function $D(k)$ is outlined.  The
decoherence funtion was derived in Ref.\cite{mckellar} as
\begin{eqnarray}
\label{eq:D}
D(k) & = & \pi \int dk'dp'dp \delta_E(k+p-k'-p')\sum_j
[V^2(\nu_{\alpha}(k),j(p)|\nu_{\alpha}(k'),j(p'))f_j(p) \nonumber \\
& + & V^2(\nu_{\alpha}(k),\overline{\nu}_{\alpha}(p)|j(k'),
\overline{j}(p'))f_{\overline{\nu}_{\alpha}}(p)],
\end{eqnarray}
where $\int dp \equiv \frac{1}{(2\pi)^3} \int d^3p$, and the $f$'s are momentum
distribution functions with
\begin{equation}
f_{\nu}^{eq}(p,\mu)=\frac{1}{1+\exp(\frac{p-\mu}{T})},
\end{equation}
and $f^{eq}_{\overline{\nu}} =  f^{eq}_{\nu}(p,\overline{\mu})$.
Note that this expression neglects Pauli blocking factors of $(1-f(p'))(1-f(k'))$.
The sum over $j$ in Eq.(\ref{eq:D}) includes all weakly interacting particle species
in the background plasma.

The matrix elements $V(i(k),j(p)|m(k'),n(p'))$ are related to weak interaction
matrix elements $M(i(k),j(p)|m(k'),n(p'))$ via
\begin{equation}
V^2(i(k),j(p)|m(k'),n(p'))=\frac{(2\pi)^3}{2k2p2k'2p'}
\delta^3(k+p-k'-p')M^2(i(k)j(p)|m(k')n(p')).
\end{equation}
The matrix elements $M(ij|mn)$ may readily be evaluated as they are simply four-Fermi
interactions, and the integration over $k'$ and $p'$ performed, to obtain
\begin{equation}
D(k)=\frac{2}{3}\frac{1}{k}\frac{G_F^2}{(2\pi)^4}
\int\frac{d^3p}{p}(k\cdot p)^2\sum_jA_jf_j(p),
\end{equation}
where the mass of the electron has been neglected, and the $A_j$'s are coefficients
which are given in terms of $\sin^2\theta_W$, and can be found, for example, in
Refs.\cite{mckellar,et}.

If we assume that all weakly interacting species are in thermal equilibrium with
zero chemical potential, then we obtain
\begin{equation}
\label{eq:D0}
D(k)=\frac{1}{2}y_{\alpha}G_F^2T^5\frac{180 \zeta(3)}{7\pi^4}\frac{k}{T},
\end{equation}
which, if we set the momentum $k$ equal to its thermal average $\langle k \rangle_0$,
reduces to the
standard expression in terms of the thermally averaged collision rate
$\langle D\rangle=\frac{1}{2}\langle\Gamma_{\alpha}\rangle$.

To allow for a nonzero lepton asymmetry, we will assume that $\nu_{\alpha}$
and $\overline{\nu}_{\alpha}$ may have non-zero chemical potential, while all other
particle species will be assumed to have zero chemical potentials.  This results
in small correction terms to Eq.(\ref{eq:D0})
\begin{equation}
D(k)=\frac{1}{2}G_F^2T^5\frac{k}{3.15T}\left\{y_{\alpha}+ u_{\alpha}\frac{\mu}{T}
+ v_{\alpha}\frac{\overline{\mu}}{T}+w_{\alpha}\left(\frac{\mu}{T}\right)^2
+ x_{\alpha}\left(\frac{\overline{\mu}}{T}\right)^2
+ O\left[\left(\frac{\mu}{T}\right)^3\right]\right\}.
\end{equation}
where $y_e \simeq 4.0$, $y_{\mu,\tau} \simeq 2.9$, $u_{\alpha} \simeq 0.72$, $v_e
\simeq 1.0$, $v_{\mu,\tau} \simeq 0.8$,
$w_{\alpha} \simeq 0.33$, $x_e \simeq 0.46$, and $x_{\mu,\tau} \simeq 0.36$.

\section{Repopulation}

We outline the approximations used in determining the form of the repopulation
function $R(k)$. The general form of this function was derived in Ref.\cite{mckellar},
and is given by
\begin{eqnarray}
\label{eq:R}
R(k) & = & \frac{2\pi}{f^{eq}(k,0)} \int dk'dp'dp \delta_E(k+p-k'-p')\times \nonumber \\
& & \sum_j [V^2(\nu_{\alpha}(k),j(p)|\nu_{\alpha}(k'),j(p'))
(f_{\nu_{\alpha}}(k')f_j(p')-f_{\nu_{\alpha}}(k)f_j(p)) \nonumber \\
& + & V^2(\nu_{\alpha}(k),\overline{\nu}_{\alpha}(p)|j(k'),\overline{j}(p'))
(f_j(k')f_{\overline{j}}(p')-f_{\nu_{\alpha}}(k)f_{\overline{\nu}_{\alpha}}(p))].
\end{eqnarray}
Note that both inelastic (annihilation) and elastic (scattering) processes are included in
the expression for $R(k)$, because both contribute to the rate at which a certain
momentum
state is refilled.  However, in the momentum averaged limit, the elastic contribution
will vanish since in that case we are only interested in the total number of particles,
and not how they are spread across the momentum distribution.\footnote{The expression given in \cite{mckellar} appears to neglect elastic processes.}

The equation for $R(k)$ is a Pauli-Boltzmann equation, and momentum states are
refilled in such a way as to drive them toward equilibrium.  The term
$(f(k')f(p')-f(k)f(p))\neq 0$ if all the $f$'s are given by equilibrium Fermi-Dirac
distributions, but this is just a consequence of neglecting the Pauli blocking factors.
The general form of $R(k)$ cannot be calculated analytically, and numerically would
require significant computing power, so it is
useful to look at approximations under which it may
be simplified.  If we assume that all the distributions are in thermal equilibrium
except for the state which is being refilled [i.e.\ $f_{\nu_{\alpha}}'(k)$], then we may
replace
$f_j(k')f_{\overline{j}}(p')$ in Eq.(\ref{eq:R}) with
$f^{eq}_{\nu_{\alpha}}(k)f^{eq}_{\overline{\nu}_{\alpha}}(p)$.  This makes sense in
terms
of pushing everything toward equilibrium, or alternatively,
\begin{eqnarray}
f^{eq}_j(k')f^{eq}_{\overline{j}}(p') & = &
f^{eq}_{\nu_{\alpha}}(k)f^{eq}_{\overline{\nu}_{\alpha}}(p)
\times \frac{(1-f^{eq}_j(k'))(1-f^{eq}_{\overline{j}}(p'))}{(1-f^{eq}_{\nu_{\alpha}}(k))
(1-f^{eq}_{\overline{\nu}_{\alpha}}(p))} \nonumber \\
& \simeq & f^{eq}_{\nu_{\alpha}}(k)f^{eq}_{\overline{\nu}_{\alpha}}(p),
\end{eqnarray}
since Pauli blocking factors are neglected.  With this approximation,
\begin{eqnarray}
\label{eq:Rv3}
R(k) & = & \frac{2\pi}{f^{eq}(k,0)}[f^{eq}_{\nu_{\alpha}}(k)-f_{\nu_{\alpha}}(k)]
\int dk'dp'dp \delta_E(k+p-k'-p') \times \nonumber \\
& &\sum_j[ V^2(\nu_{\alpha}(k),j(p)|\nu_{\alpha}(k'),j(p'))f^{eq}_j(p)
+ V^2(\nu_{\alpha}(k),\overline{\nu}_{\alpha}(p)|j(k'),\overline{j}(p'))
f^{eq}_{\overline{\nu}_{\alpha}}(p) ] \nonumber\\
& = & 2D(k)\left[\frac{f^{eq}_{\nu_{\alpha}}(k)}{f^{eq}(k,0)}
-\frac{f_{\nu_{\alpha}}(k)}{f^{eq}(k,0)}\right],
\end{eqnarray}
so that we obtain,
\begin{equation}
R(k) = \Gamma_{\alpha}(k) \left[ \frac{N^{\text{eq}}(k,\mu_{\alpha})}
{N^{\text{eq}}(k,0)}-\frac{N_{\alpha}(k)}{N^{\text{eq}}(k,0)} \right].
\label{approxR}
\end{equation}
The general expression for the repopulation function given by
Eq.(\ref{eq:R}) is
numerically intensive, so the form given by Eq.(\ref{approxR}) allows a significant
simplification to numerical calculations.

\end{document}